\documentclass[11pt]{article}
\pdfoutput=1
\usepackage{amsmath,amssymb,amsfonts,dsfont,xfrac}
\usepackage{graphicx}
\usepackage[citebordercolor={.8 .8 1},urlbordercolor ={.8 .8 1},pdfstartview=FitV]{hyperref}

\newcommand{\stress}{T}
\newcommand{\stressbar}{\bar{T}}
\newcommand{\fudge}{\sigma_{\text{\tiny{$1\!/2$}}}}

\numberwithin{equation}{section}

\begin{document}
\begin{center}

\vspace{1cm} { \Large {\bf Entanglement Entropy and Higher Spin Holography in AdS$_3$}}

\vspace{1.1cm}
Jan de Boer and Juan I. Jottar

\vspace{0.7cm}

{\it Institute for Theoretical Physics, University of Amsterdam,\\
Science Park 904, Postbus 94485, 1090 GL Amsterdam, The Netherlands}

{\tt J.deBoer@uva.nl, J.I.Jottar@uva.nl} \\

\vspace{1.5cm}

\end{center}

\begin{abstract}
\noindent A holographic correspondence has been recently developed between higher spin theories in three-dimensional anti-de Sitter space (AdS$_3$) and two-dimensional Conformal Field Theories (CFTs) with extended symmetries. A class of such dualities involves $SL(N,\mathds{R})\times SL(N,\mathds{R})$ Chern-Simons gauge theories in the (2+1)-dimensional bulk spacetime, and CFTs with $\mathcal{W}_N$ symmetry algebras on the (1+1)-dimensional boundary. The topological character of the Chern-Simons theory forces one to reconsider standard geometric notions such as black hole horizons and entropy, as well as the usual holographic dictionary. Motivated by this challenge, in this note we present a proposal to compute entanglement entropy in the  $\mathcal{W}_N$ CFTs via holographic methods. In particular, we introduce a functional constructed from Wilson lines in the bulk Chern-Simons theory that captures the entanglement entropy in the CFTs dual to standard AdS$_3$ gravity, corresponding to $SL(2,\mathds{R})\times SL(2,\mathds{R})$ gauge group, and admits an immediate generalization to the higher spin case. We explicitly evaluate this functional for several known solutions of the Chern-Simons theory, including charged black holes dual to thermal CFT states carrying higher spin charge, and show that it reproduces expected features of entanglement entropy, study whether it obeys strong subadditivity, and moreover show that it 
reduces to the thermal entropy in the appropriate limit.

%\vspace{30pt}
%\centerline{Draft version of \today}
\end{abstract}

\pagebreak

\setcounter{page}{1}
\setcounter{equation}{0}
\tableofcontents

\pagebreak

%%%%%%%%%%%%%%%%%%%%%%%%%%%%%%%%%%%%%%%%%%%%%%%%%%%%%%%%%%%%%%%%%%%%%%%%%%%%%%%%
\section{Introduction \label{sec: Intro}}
%%%%%%%%%%%%%%%%%%%%%%%%%%%%%%%%%%%%%%%%%%%%%%%%%%%%%%%%%%%%%%%%%%%%%%%%%%%%%%%%
Over the last two decades, the holographic principle \cite{tHooft:1993gx,Susskind:1994vu} has become a cornerstone of theoretical physics. Put simply, it asserts that certain theories of gravity in $(d+1)$ dimensions can be described as quantum field theories in $d$ dimensions, and vice versa. To date, the most successful and concrete realization of holography is the Anti-de Sitter (AdS)/Conformal Field Theory (CFT) correspondence of \cite{Maldacena:1997re} (see \cite{Witten:1998qj,Gubser:1998bc} also), which proposes a duality between conformally-invariant gauge theories, and string theory on a space in one higher dimension that asymptotes to AdS, the maximally-symmetric solution of Einstein's equations with negative cosmological constant. Notably, the ideas behind the AdS/CFT correspondence have shed light on the structure of quantum gravity and gauge theories alike, while providing powerful computational techniques to explore the connections between them. 

The holographic correspondence is an example of strong-weak duality; so far, the vast majority of the work in AdS/CFT has focused on the regime where the gauge theory is strongly-coupled and the string theory description becomes weakly-coupled, effectively reducing to classical (super)gravity. To a high extent, the power of the correspondence lies in the fact that  many problems can be addressed analytically in the latter theory, providing insight into the regime where the quantum field theory becomes intractable with the standard perturbative techniques. Quite remarkably, in light of this feature holographic techniques have found a niche of applications in condensed matter physics, where strongly-correlated systems are routinely engineered and studied in the laboratory. It is however of considerable theoretical importance to understand and test the holographic duality in other regimes of couplings as well. A particularly interesting example which is outside the scope of the usual AdS/CFT correspondence is the conjecture \cite{Klebanov:2002ja} of Klebanov and Polyakov relating critical $O(N)$ vector models in the large-$N$ limit to the higher spin Fradkin-Vasiliev theory in AdS$_{4}$ \cite{Fradkin:1987ks,Fradkin:1986qy}, for which a considerable amount of evidence has been provided recently (see \cite{Giombi:2012ms} and references therein). 

Although of a somewhat different flavor, similar dualities have been put forward in lower dimensions, an interesting example being the proposal of \cite{Gaberdiel:2010pz} relating the three-dimensional Vasiliev higher spin theory and the large-$N$ limit of $\mathcal{W}_{N}$ minimal coset CFTs. An aspect that makes the lower-dimensional setup particularly appealing is the fact that universal results for two-dimensional CFTs, such as the Cardy entropy formula \cite{Cardy:1986ie,Bloete:1986qm} and the zero- and finite-temperature entanglement entropy in $1d$ systems \cite{Holzhey:1994we,Calabrese:2004eu}, are known to be recovered in the framework of the standard AdS$_{3}$/CFT$_{2}$ correspondence. In particular, the thermal entropy of the three-dimensional (BTZ) black hole as computed with the standard Bekenstein-Hawking formula precisely matches the form predicted by Cardy's asymptotic growth of states in a unitarity CFT (see e.g. \cite{Strominger:1997eq,Carlip:2005zn,Kraus:2006wn}). Similarly, the single-interval entanglement entropy of the CFT state dual to this black hole is reproduced using the Ryu-Takayanagi (R-T) prescription to compute entanglement entropy holographically \cite{Ryu:2006bv,Ryu:2006ef}. 
Recently, some universal aspects of the AdS$_{3}$/CFT$_{2}$ correspondence were further elucidated in \cite{Hartman:2013mia,Faulkner:2013yia}.
An interesting question is then whether (and how) this universality prevails in the presence of extended symmetries furnished by higher spin operators.

 As an added bonus, given that the higher-dimensional theories of interacting massless higher spin fields are technically involved and difficult to work with, it is desirable to work with models that retain their key features while being amenable to study, such as the AdS$_{3}$ higher spin theories. In fact, in three dimensions it is possible to truncate the tower of massless modes to retain fields of spin $s \leq N$ only \cite{Blencowe:1988gj}. This is to be contrasted with their higher-dimensional counterparts of the Fradkin-Vasiliev type, where an infinite number of higher spin fields must be kept. Furthermore, the corresponding higher spin theories in AdS$_{3}$ can be cast in the form of an $SL(N,\mathds{R})\times SL(N,\mathds{R})$ Chern-Simons gauge theory, and many of the familiar techniques to analyze such theories can be brought to bear. 

Recently, many entries of the holographic dictionary for higher spin AdS$_{3}$ theories have been established. In particular, much in the same way that standard Einstein gravity with AdS$_{3}$ boundary conditions has an asymptotic symmetry group generated by two copies of the Virasoro algebra acting on the spacetime boundary \cite{Brown:1986nw}, the analysis of asymptotic symmetries in the higher spin case \cite{Henneaux:2010xg,Campoleoni:2010zq,Gaberdiel:2010ar} has shown them to correspond to two-dimensional CFTs with extended symmetry algebras of (classical) $\mathcal{W}_{N}$ type, in agreement with earlier expectations \cite{deBoer:1998ip}. In contrast with the universal results quoted above, not much is known about the CFTs in the presence of deformations by higher spin operators, making the results obtained from holography all the more interesting. Motivated by these facts, in the present paper we initiate the study of entanglement entropy in higher spin holography in AdS$_{3}\,$.

 Perhaps the main challenge in extending the usual AdS$_{3}$ holographic dictionary to the higher spin case is that we must surrender the traditional geometric interpretation of notions such as black hole horizons and entropy, which lie at the core of AdS/CFT dualities in higher dimensions, and formulate them in a language which is appropriate in light of the topological character of the bulk theory. Indeed, the standard R-T prescription to compute entanglement entropy holographically is intrinsically geometric in nature: in order to obtain the entanglement entropy of a region $A$ in the boundary theory, one is instructed to find the minimal area bulk surface that is anchored on the boundary of $A$ and ``dips" into the bulk spacetime. In order to solve this problem, in the present work we introduce a functional that is naturally defined in terms of Wilson lines in the bulk Chern-Simons theory and captures the entanglement entropy in the situations where an independent field-theoretical result is available (namely in the absence of higher spin charges), while generalizing naturally to the higher spin setup. In the absence of explicit entanglement entropy results for CFTs perturbed by higher spin operators, we apply our proposal in several examples and show that it reproduces the properties that the field theory entanglement is expected to satisfy, such as strong subadditivity (up to some subtleties). Moreover, building on our previous general results for the thermal entropy in higher spin theories \cite{deBoer:2013gz}, we will show that our entanglement functional is constructed so that it approaches the thermal entropy in the high temperature limit in which the entanglement becomes extensive, even in the presence of non-trivial higher spin charges and chemical potentials.

The structure of the paper is as follows. In section \ref{sec:CS gravity} we briefly review the formulation of standard Einstein gravity in three dimensions as a Chern-Simons theory and its extension to include a finite number of higher spin fields, regarded as $SL(N,\mathds{R})\times SL(N,\mathds{R})$ Chern-Simons theory (with $N > 2$).  In section \ref{sec: proposal} we introduce our proposed bulk functional and explicitly show that it reproduces the known CFT entanglement entropy in the absence of higher spins (namely in the $SL(2,\mathds{R})\times SL(2,\mathds{R})$ case corresponding to standard Einstein gravity). We then discuss various properties of this functional in the general case, and in particular apply our previous results \cite{deBoer:2013gz} for the thermodynamic entropy of higher spin black holes to show how to choose the representation of the algebra in which the Wilson lines are evaluated, in such a way that our entanglement functional also reproduces the thermal entropy of the dual CFTs in the appropriate limit. In section \ref{sec: sl3 examples} we apply our proposal to some of the known solutions of the spin-3 theory, including the charged black hole solution of \cite{Gutperle:2011kf} which is dual to a CFT ensemble at finite temperature and finite higher spin charge. We conclude in section \ref{sec: conclusions} with a discussion of our results and outlook. The details of our conventions and some useful facts and calculations are collected in the appendices. 

\noindent \textbf{Note: } While this work was being completed we became aware of \cite{Castro:2013}, where a similar proposal to compute entanglement entropy in three-dimensional higher spin theories will be put forward. Their formulation is based on the observation that, for $N=2$, the geodesic distance on AdS$_{3}$ interpreted as a group manifold can be computed in terms of Wilson lines in an infinite-dimensional representation of the gauge group.

%%%%%%%%%%%%%%%%%%%%%%%%%%%%%%%%%%%%%%%%%%%%%%%%%%%%%%%%%%%%%%%%%%%%%%%%%%%%%%%%
\section{Higher spin theories in $\text{AdS}_{3}$}\label{sec:CS gravity}
%%%%%%%%%%%%%%%%%%%%%%%%%%%%%%%%%%%%%%%%%%%%%%%%%%%%%%%%%%%%%%%%%%%%%%%%%%%%%%%%
As it is well-known, three-dimensional gravity with a negative cosmological constant can be formulated as a Chern-Simons theory \cite{Achucarro:1987vz,Witten:1988hc} (see \cite{Witten:2007kt} for a modern perspective). The extension to higher spin theories utilizes the Chern-Simons language, and in fact resembles the pure gravity case in many a way. Therefore, we begin with a brief discussion of standard gravity with AdS boundary conditions in the Chern-Simons formulation. Our conventions and some extra details can be found in appendix \ref{sec: vielbein formalism}. 

%%%%%%%%%%%%%%%%%%%%%%%%%%%%%%%%%%%%%%%%%%%%%%%
\subsection{AdS$_{3}$ gravity as a Chern-Simons theory}
%%%%%%%%%%%%%%%%%%%%%%%%%%%%%%%%%%%%%%%%%%%%%%%
Let $a,b,\ldots$ denote local Lorentz indices in $(2+1)$ dimensions, and define the dual $\omega^{a}$ of the spin connection as 
\begin{equation}\label{definition dual spin connection}
\omega^{a} \equiv \frac{1}{2}\epsilon^{abc}\omega_{bc}
\end{equation}

\noindent or equivalently $\omega_{ab} = -\epsilon_{abc}\,\omega^{c}$. We can then combine this object with the dreibein or ``triad" $e^{a}$ into $so(2,1)\simeq sl(2,\mathds{R})$ connections $A$, $\bar{A}$ defined as
\begin{equation}\label{def CS connecions}
A = \omega +\frac{e}{\ell}\,,\qquad  \bar{A}=  \omega - \frac{e}{\ell}\,,
\end{equation}
\noindent where $\ell$ is the AdS$_{3}$ radius, namely the length scale set by the cosmological constant ($\Lambda_{cosmo} = -1/\ell^2$). Here, $\omega \equiv \omega^{a}J_{a}$ and $e \equiv e^{a}J_{a}\,$, and the generators $J_{a}$ obey the $so(2,1)\simeq sl(2,\mathds{R})$ algebra $\left[J_{a},J_{b}\right] = \epsilon_{abc}\,\eta^{cd}J_{d} = \epsilon_{ab}^{\hphantom{ab}c}J_{c}\,
$ (the relation between the $so(2,1)$ generators and the $sl(2,\mathds{R})$ generators $\Lambda^{0}$, $\Lambda^{\pm}$ is given in \eqref{def sl2 generators}). We emphasize that the bar notation \textit{does not} denote complex conjugation in Lorentzian signature.

Defining the Chern-Simons form $CS$ as
\begin{equation}
CS(A) = A\wedge dA + \frac{2}{3}A\wedge A\wedge A
\end{equation}
\noindent  one finds that the combination $\mbox{Tr}\Bigl[CS(A) - CS(\bar{A})\Bigr]$ yields the Einstein-Hilbert Lagrangian, up to a total derivative. The precise relation is (c.f. \eqref{Chern Simons action})
\begin{align}\label{CS action}
I \equiv{}&
 \frac{k}{4\pi }\int_{M} \mbox{Tr}\Bigl[CS(A) - CS(\bar{A})\Bigr]
\\
={}&
 \frac{1}{16\pi G_{3}}\left[\int_{M} d^{3}x\sqrt{|g|}\left(\mathcal{R} + \frac{2}{\ell^{2}}\right) -\int_{\partial M}\omega^{a}\wedge e_{a}\right],
 \nonumber
\end{align}
\noindent  where $G_{3}$ is the $3d$ Newton constant. When evaluated on shell, the boundary term amounts to $(1/2)$ times the standard Gibbons-Hawking surface term. Normalizing the $so(2,1)$ generators according to $\mbox{Tr}\left[J_{a}J_{b}\right] = \eta_{ab}/2\,$, we identify the Chern-Simons level $k$ as
\begin{equation}\label{sl2 k}
k = \frac{\ell}{4G_{3}}\,.
\end{equation}
\noindent Whether the gauge group is $SO(2,1)\times SO(2,1)$ or some locally isomorphic (but globally inequivalent) cover is a question that has consequences for the quantization of $k$, and affects the values of the central charges in the dual field theory (see \cite{Witten:2007kt,Maloney:2007ud}, for example). 

One can easily establish a dictionary between the standard (metric) and Chern-Simons formulations of the theory.  For example, the metric tensor is obtained from the triad as $g_{\mu\nu} = 2\text{Tr}\left[e_{\mu}e_{\nu}\right]\,$, and Einstein's equations translate into the flatness of the gauge connections,
\begin{align}
F = dA + A\wedge A =0\,,
\qquad \bar{F} = d\bar{A} + \bar{A}\wedge \bar{A} = 0\,.
\end{align}
\noindent In components,  $F_{\mu\nu} = \partial_{\mu}A_{\nu} - \partial_{\nu}A_{\mu} + \left[A_{\mu},A_{\nu}\right]=0$ as usual. Similarly, under an infinitesimal gauge transformation $\delta A = d\lambda +\left[A,\lambda\right]\,$, $\delta\bar{A} = d\bar{\lambda} + \left[\bar{A},\bar{\lambda}\right]\,$, the dreibein transforms as
\begin{equation}\label{transformation vielbein}
\delta e_{\mu} = e_{\nu}\,\xi^{\nu}_{\hphantom{\nu};\mu} + \frac{1}{2}\left[e_{\mu},\lambda + \bar{\lambda}\right],
\end{equation}
\noindent where the infinitesimal generator $\xi^{\mu}$ is defined in terms of the inverse triad as $\xi^{\mu} = (\ell/2)e_{a}^{\hphantom{a}\mu}\left(\lambda^{a}- \bar{\lambda}^{a}\right)$. The first term in \eqref{transformation vielbein} gives rise to the standard infinitesimal diffeomorphisms acting on the metric, while the second term represents a rotation of the local Lorentz frame.

That one can rephrase three-dimensional gravity as a topological theory is a reflection of the fact that the dynamical degrees of freedom in the theory are not local: as it is well-known, all solutions of the three-dimensional Einstein's equations with negative cosmological constant are locally equivalent to AdS$_{3}\,$. The non-triviality of the dynamics is rooted in the existence of globally inequivalent solutions, such as black holes, and boundary excitations. Naturally, the latter are intimately tied to the choice of boundary conditions, which are a crucial ingredient in holographic constructions. As first shown by Brown and Henneaux \cite{Brown:1986nw}, in standard three-dimensional gravity with negative cosmological constant one can choose consistent boundary conditions such that the asymptotic symmetries correspond to two copies of the Virasoro algebra with central charge $c =  6k = 3\ell/(2G_{3})\,$. 

Let us briefly review how the Brown-Henneaux result comes about in the Chern-Simons formulation, as first derived by \cite{Coussaert:1995zp}. We consider Chern-Simons theory on a Lorentzian three-dimensional manifold $M$ with topology $\mathds{R}\times D\,$, where the $\mathds{R}$ factor corresponds to the timelike direction and $D$ is a two-dimensional manifold with boundary $\partial D \simeq S^{1}$. We will introduce coordinates $(\rho,t,\varphi)$ on $M$, where $\rho$ is the bulk radial coordinate and the constant-$\rho$ surfaces (in particular the asymptotic boundary $\partial M$ at $\rho \to \infty$) have the topology of a cylinder.  Given a set of boundary conditions, the \textit{asymptotic symmetry algebra} is defined as the set of transformations (diffeomorphisms in this case) that respect the boundary conditions, modulo trivial gauge transformations which are generated by constraints. The charges associated with the asymptotic symmetries generate global transformations that take us between distinguishable physical states in phase space (which becomes a Hilbert space upon quantization). Imposing boundary conditions $\left.A_{-}\right|_{\partial M} \to 0$, $\left.\bar{A}_{+}\right|_{\partial M} \to 0\,$, one finds that the asymptotic symmetries correspond to two copies of an affine $sl(2,\mathds{R})$ algebra at level $k\,$. Equivalently, at this stage the Chern-Simons theory plus boundary conditions becomes a non-chiral Wess-Zumino-Witten (WZW) model.  Further imposing that the connection approaches an AdS$_{3}$ connection at the boundary, $A - A_{AdS_{3}} \xrightarrow[\rho \to \infty]{} \mathcal{O}(1)\,$, the asymptotic symmetries reduce to two copies of the Virasoro algebra with central charge $c = 6k = 3\ell/(2G_{3})\,$. This is an example of the so-called Drinfeld-Sokolov reduction \cite{Drinfeld:1984qv}: denoting the modes of the Kac-Moody currents by $J^{a}_{n}\,$, the requirement of AdS asymptotics translates to $J^{0}_{n} = 0\,, J^{+}_{n} \simeq k\delta^{0}_{n}\,$, reducing the current algebra to the Virasoro symmetries.

We will now review some solutions that will play an important role later on. As we mentioned above, all the solutions of three-dimensional Einstein gravity with negative cosmological constant are locally connected to AdS$_{3}$ by a change of coordinates. In \cite{Banados:1998gg} it was pointed out that the metric
\begin{equation}\label{general AAdS3 metric}
ds^{2} = \ell^{2}\left[d\rho^{2} +\frac{T(x^+)}{k}\,(dx^{+})^{2} + \frac{\bar{T}(x^-)}{k}\,(dx^{-})^2- \left(e^{2\rho} +\frac{T(x^+)\bar{T}(x^-)}{k^{2}} e^{-2\rho}\right)dx^{+}dx^{-} \right],
\end{equation}
\noindent where $x^{\pm} = t/\ell \pm \varphi\,$, is a solution of Einstein's equations for any functions $T = T(x^{+})$, $\bar{T} = \bar{T}(x^{-})$. Furthermore, it represents the whole space of asymptotically AdS$_{3}$ (AAdS$_{3}$) solutions with a flat boundary metric at $\rho \to \infty$. In particular, the BTZ black hole \cite{Banados:1992wn} with mass $M$ and angular momentum $J$ \footnote{One restricts $|J| \leq \ell M$ in order to avoid naked singularities. The value $|J| = \ell M$ that saturates the bound corresponds to the extremal (zero temperature) BTZ black hole.} is obtained for constant $\stress$, $\stressbar$ given as
\begin{equation}
\begin{aligned}\label{BTZ L and Lbar}
\stress_{BTZ} &= \frac{1}{2}\left(M\ell  - J\right) =  k\frac{\pi^2\ell^{2}}{\beta_{-}^{2}}\,
\\
\stressbar_{BTZ} &= \frac{1}{2}\left(M\ell  + J\right) = k\frac{\pi^2\ell^{2}}{\beta_{+}^{2}}\, ,
\end{aligned}
\end{equation}
\noindent  where we introduced the inverse chiral temperatures $\beta_{\pm} = 1/T_{\pm}$. Via the standard holographic dictionary, the functions $T$ and $\bar{T}$ are seen to correspond to the stress tensor modes in the dual CFT (see \cite{Kraus:2006wn} for a review of the AdS$_{3}$/CFT$_{2}$ correspondence); in particular, the zero modes of $T$ and $\bar{T}$ are the eigenvalues of the operators $L_0$ and $\bar{L}_0\,$. In this parameterization the global AdS$_{3}$ solution corresponds to $J=0$ and $M\ell = -k/2$ ($8GM = -1$), i.e.
\begin{align}\label{AdS3 L and Lbar}
\stress_{AdS_{3}} &= \stressbar_{AdS_{3}}=-\frac{k}{4}\,,
\end{align}

\noindent while the so-called Poincar\'e patch of AdS$_{3}$ is obtained with $J=M=0\,$ (provided we un-compactify the  boundary spatial coordinate). 

We will choose a basis of generators $\{\Lambda^{0},\Lambda^{\pm}\}$ for the  $sl(2,\mathds{R})$ algebra, satisfying
\begin{equation}
\left[\Lambda^{\pm},\Lambda^{0}\right]=\pm \Lambda^{\pm}\,,\quad \left[\Lambda^{+},\Lambda^{-}\right]=2\Lambda^{0}\,.
\end{equation} 

\noindent In order to write down the above solutions in the Chern-Simons formulation, one first notices that the gauge freedom allows one to fix the radial dependence as
\begin{equation}
\begin{aligned}\label{general form connections}
A
={}&
 b^{-1} db + b^{-1}a(x^+,x^-)\,b
 \\
\bar{A} 
={}&
 b\,db^{-1} +b\,\bar{a}(x^+,x^-)\,b^{-1}\, ,
\end{aligned}
\end{equation}
\noindent with $b = b(\rho) = e^{\rho \Lambda^{0}}\,$.  The $\rho$-independent connections $a$, $\bar{a}$ corresponding to \eqref{general AAdS3 metric} are then given by
\begin{equation}
\begin{aligned}\label{sl2r connections}
a
&=
 \left(\Lambda^+ - \frac{\stress(x^{+})}{k}\Lambda^{-}\right)dx^{+}\,,
\\
 \bar{a}
 &=
  \left(-\Lambda^{-}+ \frac{\stressbar(x^{-})}{k} \Lambda^{+}\right)dx^{-}\,.
  \end{aligned}
\end{equation}

Since Chern-Simons theory is a theory of flat connections, we can \textit{locally} write its solutions in terms of group elements $g$, $\bar{g}$ as follows:
\begin{align}\label{BTZ connections}
A
=
 g^{-1} dg\,,\qquad  \bar{A} = \bar{g}^{-1} d\bar{g}\,.
\end{align}
\noindent For example, for the above solutions with constant $\stress$, $\stressbar$ (which include globally-defined black holes) we find
\begin{equation}
\begin{aligned}\label{BTZ group elements}
g 
&=
 \exp\left\{\left[\Lambda^+ - \left(\frac{\stress}{k}\right)\Lambda^{-}\right]x^{+}\right\}b(\rho)\,,&
\\
\bar{g}
&=
\exp\left\{-\left[\Lambda^{-} - \left(\frac{\stressbar}{k}\right) \Lambda^{+}\right]x^{-}\right\}b^{-1}(\rho)\,.&
\end{aligned}
\end{equation}
\noindent Additional care must be exercised in the presence of non-contractible cycles, such as the $\varphi$ circle parameterizing the horizon in three-dimensional black hole geometries: if the connection has non-trivial holonomy, it undergoes a gauge transformation upon transport around the horizon. In other words, the group elements $g$, $\bar{g}$ in \eqref{BTZ connections} are not, in general, single-valued. This will be important for us below when we discuss how to recover the thermodynamic entropy in the limit in which the entanglement entropy becomes extensive.

%%%%%%%%%%%%%%%%%%%%%%%%%%%%%%%%%%%%%%%%%%%%%%%
\subsection{The $SL(N,\mathds{R})\times SL(N,\mathds{R})$ higher spin theory}
%%%%%%%%%%%%%%%%%%%%%%%%%%%%%%%%%%%%%%%%%%%%%%%
Having rephrased the standard AdS$_{3}$ Einstein gravity as an $SL(2,\mathds{R})\times SL(2,\mathds{R})$ Chern-Simons gauge theory, we now introduce higher spins by promoting the gauge group to $SL(N,\mathds{R})\times SL(N,\mathds{R})\,$. When $N > 2\,$, this theory describes gravity coupled to a tower of fields of spin $s \leq N$ \cite{Blencowe:1988gj}. The precise field content of the gravitational theory (and hence the spectrum and symmetry algebra of the dual CFT) depends on how the $sl(2,\mathds{R})$ subalgebra associated to the gravity sector is embedded into $sl(N,\mathds{R})\,$ (see \cite{Bais:1990bs,deBoer:1993iz}). The different embeddings are characterized by the way in which the fundamental representation of $sl(N,\mathds{R})$ decomposes into $sl(2,\mathds{R})$ representations, and these branching rules are in turn classified by integer partitions of $N\,$. As a concrete example, consider the defining representation $\mathbf{3}_{3}$ of $sl(3,\mathds{R})$. Denoting the $(2j +1)$-dimensional representation of $sl(2,\mathds{R})$ by $\mathbf{(2j+1)}_{2}\,$, the non-trivial inequivalent embeddings are characterized by the branching rules $\mathbf{3}_{3} \to \mathbf{3}_{2}\,$ and $\mathbf{3}_{3} \to \mathbf{2}_{2}\oplus \mathbf{1}_{2}\,$. The first embedding is the so-called ``principal embedding", characterized by the fact that the fundamental representation becomes an irreducible representation of the embedded algebra. The second embedding is called ``diagonal embedding", because the embedded $sl(2,\mathds{R})$ takes a block-diagonal form inside $sl(3,\mathds{R})\,$. 

The branching of the $\left(N^2-1\right)$-dimensional adjoint representation can be determined from that of the fundamental representation, and one deduces in this way the decomposition of the algebra itself and hence the spectrum \cite{Bais:1990bs,deBoer:1992sy}. In the principal embedding, $\text{adj}_N \to \mathbf{3}_{2} \oplus \mathbf{5}_{2} \oplus \ldots \oplus \mathbf{(2N-1)}_{2}\,$ showing that the $sl(N,\mathds{R})$ algebra decomposes into $N-1$ representations with $sl(2,\mathds{R})$ spins ranging from $1$ to $N-1\,$ (the spin 1 multiplet being the $sl(2,\mathds{R})$ generators themselves). From the perspective of the bulk theory, these representations correspond to the metric ($g_{\mu\nu} \sim \text{Tr}[e_{\mu}e_{\nu}]$) and a tower of symmetric tensor fields with spins $3,\ldots,N\,$ ($\phi_{\mu\nu\rho} \sim \text{Tr}[e_{(\mu}e_{\nu}e_{\rho)}]$ and so forth). In general, the conformal weight of the corresponding operators in the boundary theory is obtained by adding one to the $sl(2,\mathds{R})$ spin (see \cite{Ammon:2011nk} for example). Consequently, in addition to the stress tensor, in the principal embedding one finds primary operators of weight $3,4\ldots, N\,$. In the diagonal embedding one has instead $\text{adj}_{N} \to \mathbf{3}_2 \oplus 2(N-2)\cdot \mathbf{2}_2 \oplus (N-2)^{2} \cdot \mathbf{1}_2\,$. Hence, the spectrum in the diagonal embedding contains currents of weight $1$ and $3/2\,$. When charged fields are present, there is always a consistent truncation where they are taken to be zero; in the diagonal embedding, this corresponds to setting the spin $3/2$ fields to zero while truncating the weight-one currents to the diagonal subset. From the bulk perspective, the theory in the diagonal embedding then contains a truncation to standard gravity coupled to $U(1)^{2(N-2)}$ gauge fields. Indeed, as discussed in \cite{Castro:2011fm}, a class of black hole solutions in the diagonal embedding correspond to BTZ black holes charged under Abelian holonomies.

The asymptotic symmetry analysis in the $N> 2$ case  was performed in \cite{Henneaux:2010xg,Campoleoni:2010zq} (see \cite{deBoer:1991jc,DeBoer:1992vm} for early work), and parallels the $N=2$ discussion in \cite{Coussaert:1995zp} closely. In particular one imposes ``Drinfeld-Sokolov boundary conditions" as before,
\begin{equation}\label{AAdS3 bcs}
A_{-}=0\,,\qquad A_{\rho} = b^{-1}(\rho)\partial_{\rho}b(\rho)\,,\qquad
 A - A_{AdS_{3}} \xrightarrow[\rho \to \infty]{\phantom{\rho \to \infty}} \mathcal{O}(1)\,,
\end{equation}

\noindent and similarly for the barred connection (notice that the last condition on $A$ does not imply that $a$ obeys that condition as well, and that \eqref{sl2r connections} is compatible with \eqref{AAdS3 bcs}). The asymptotic symmetries are then given by two copies of the so-called $\mathcal{W}_{N}$ algebras \cite{Zamolodchikov:1985wn}, which correspond to non-linear extensions of the Virasoro algebra. As a concrete example, for the $SL(3,\mathds{R})\times SL(3,\mathds{R})$ theory in the principal embedding the corresponding asymptotic symmetry algebra consists of two copies of the $\mathcal{W}_{3}$ algebra, with classical 
central charge $c = 6k = 3\ell/(2G_{3})\,$. According to the general features discussed above, the algebra in this case includes the stress tensor and primary operators of weights $(3,0)$ and $(0,3)$. For $N=3$ there is only one other non-trivial inequivalent embedding, i.e. the diagonal embedding. The asymptotic symmetry algebra in this case is identified with the so-called $\mathcal{W}_{3}^{(2)}$ algebra \cite{Henneaux:2010xg,Campoleoni:2010zq,Ammon:2012wc}. Besides the stress tensor, this algebra contains two weight-$3/2$ primary operators and a weight one current, with classical central charge given by $\hat{c} = c/4 = 3k/2\,$. Different boundary conditions giving rise to non-AAdS$_{3}$ higher spin theories have been also considered recently in \cite{Gary:2012ms,Afshar:2012nk}.
 
Let us write the coefficient of the Chern-Simons action in the higher spin case as $k_{cs}/(4\pi)\,$. Since the trace in the action is taken in the fundamental representation, matching with the normalization of the Einstein-Hilbert action requires
\begin{equation}\label{kcs}
k_{cs} = \frac{k}{2\mbox{Tr}_N\left[\Lambda^0\Lambda^0\right]}
\end{equation}

\noindent where, according to the above discussion, $k$ is the level of the $sl(2,\mathds{R})\times sl(2,\mathds{R})$ Chern-Simons theory contained in the full theory through the choice of embedding, and $\text{Tr}_{N}$ denotes the trace in the fundamental ($N$-dimensional) representation. In terms of the level $k_{cs}$ of the $sl(N,\mathds{R})\times sl(N,\mathds{R})$ theory, the central charge in the boundary CFT is given by
\begin{equation}\label{central charge}
c  = 12k_{cs}\text{Tr}_N\left[\Lambda^0\Lambda^0\right].
\end{equation}

\noindent Note that, for fixed $k_{cs}\,$, the central charge will be different for different embeddings.

%%%%%%%%%%%%%%%%%%%%%%%%%%%%%%%%%%%%%%%%%%%%%%%%%%%%%%%%%%%%%%%%%%%%%%%%%%%%%%%%
\section{A holographic entanglement entropy proposal for higher spin theories}\label{sec: proposal}
%%%%%%%%%%%%%%%%%%%%%%%%%%%%%%%%%%%%%%%%%%%%%%%%%%%%%%%%%%%%%%%%%%%%%%%%%%%%%%%%
Consider a quantum system described by a density matrix $\rho\,$, and divide it into two subsystems $A$ and $B = A^{c}\,$. The \textit{reduced density matrix} $\rho_{A}$ of subsystem $A$ is defined by tracing over the degrees of freedom in $B\,$, i.e. $\rho_{A}=\text{Tr}_{B}\,\rho\,$. The \textit{entanglement entropy} $S_{A}$ of $A$ is then defined as the von Neumann entropy associated with $\rho_{A}\,$:
\begin{equation}
S_{A} = -\text{Tr}_{A}\,\rho_{A}\log \rho_{A}\,.
\end{equation}

\noindent If the full system was originally in a pure state, i.e. $\rho = |\Psi \rangle \langle \Psi|\,$, then $S_{A} = S_{B}\,$. This property does not hold if the system was originally in a mixed state, such as a thermal ensemble with density matrix $\rho=e^{-\beta H}\,$. 

From a theoretical standpoint, the entanglement entropy has several interesting properties associated with its non-local nature, and can moreover serve as a useful tool to characterize gapped phases of matter in the absence of classical order parameters and spontaneous symmetry breaking \cite{Kitaev:2005dm,Levin:2006zz}. Unfortunately, field-theoretical calculations of entanglement entropy are in general notoriously difficult to perform, even for free theories. In theories with a (standard) gravity dual, however, entanglement entropies can be computed in a rather straightforward manner using an elegant holographic prescription due to Ryu and Takayanagi \cite{Ryu:2006bv,Ryu:2006ef}  (see \cite{Nishioka:2009un} for a review). Suppose we want to compute the entanglement entropy associated with a spatial region $A$ in the field theory. The R-T recipe instructs us to construct the minimal spacelike surface $\gamma_{A}$ that is anchored at the boundary $\partial A$ of $A$ and extends into the bulk spacetime. Then, the corresponding entanglement entropy is obtained in terms of the area of $\gamma_{A}$ as $S_{A} = \text{Area}(\gamma_{A})/(4G)\,$, where  $G$ is the Newton constant associated to the bulk spacetime. The prescription correctly reproduces the area law of entanglement entropy, and it has been shown to be strongly-subadditive \cite{Headrick:2007km} as well. Moreover, it has been generalized to include cases where the field theory state is time-dependent \cite{Hubeny:2007xt}. Strong evidence for the correctness of the R-T prescription has been given in \cite{Fursaev:2006ih,Lewkowycz:2013nqa}.

In the present context we will focus on situations where the full system is a $(1+1)$-dimensional CFT, and consider subsystems determined by spatial (equal time) intervals. 
Due to the large amount of symmetry that $(1+1)$-dimensional CFTs enjoy, a variety of quantities can be computed in closed form.
Indeed, using CFT techniques, universal results have been derived for the single-interval entanglement entropy at zero and finite temperature \cite{Holzhey:1994we,Calabrese:2004eu}. From the point of view of holography, in the particular case of a three-dimensional bulk and a two-dimensional boundary theory the minimal surface prescription of R-T amounts to finding the length of a geodesic in an asymptotically AdS$_{3}$ (AAdS$_{3}$) spacetime, and it correctly reproduces the known field theory results. On the other hand, to the extent of our knowledge there are no analytic results for entanglement entropy in the presence of deformations by higher spin currents, or in states carrying non-trivial higher spin charges, and we would therefore like to extend the holographic calculations to encompass these situations.

Since the bulk theory under consideration is topological, a reasonable starting point is to rephrase the geometric statement of the R-T proposal in terms of the natural building blocks at our disposal in the gauge theory, such as Wilson lines. To this end, given two points $P$ and $Q$ in the bulk spacetime, we start by considering the following ``composite" Wilson loop
\begin{align}\label{definition composite loop}
 W_{\mathcal{R}}(P,Q)
\equiv{}&
\mbox{Tr}_{\mathcal{R}}\Biggl[\mathcal{P}\exp\left(\int_{Q}^{P}\bar{A}\right)\,\mathcal{P}\exp\left(\int_{P}^{Q}A\right)\Biggr] 
\end{align}

\noindent where $\mathcal{P}$ denotes the usual path ordering, and the trace is evaluated in a representation that will be specified later on. As we have discussed, the gauge connections undergo a gauge transformation upon transport around a cycle with non-trivial holonomy; locally, however, we can write the flat connections as in \eqref{BTZ connections}, and $W(P,Q)$ reduces to
\begin{align}\label{definition composite loop 2}
 W_{\mathcal{R}}(P,Q)
=
 \mbox{Tr}_{\mathcal{R}}\Bigl[\bar{g}^{-1}(P)\bar{g}(Q)g^{-1}(Q)g(P)\Bigr].
\end{align}

\noindent Up to global issues (such as winding around a non-contractible cycle), we see that the result is path-independent, i.e. it depends on the positions of the endpoints $P$ and $Q$ only. One may worry about the lack of obvious gauge invariance of this expression, but as we explain below this is no
cause for concern. One may also worry that different, homotopically inequivalent paths may exist that connect $P$ and $Q$, on which the answer
clearly depends. We will address this issue below as well.

To gain some intuition about the significance of the functional \eqref{definition composite loop}, we first evaluate it for the AAdS$_{3}$ solutions of the  $SL(2,\mathds{R})\times SL(2,\mathds{R})$ theory. Plugging in the solutions \eqref{BTZ group elements} with constant $\stress$, $\stressbar$ and taking the trace in the 2$d$ (defining) representation \eqref{2d representation of sl2r} of $sl(2,\mathds{R})$ we obtain
\begin{align}\label{loop for AAdS3}
W_{2d}&(P,Q)
=
 2\cosh(\rho_{P}-\rho_{Q}) \cosh\left[\sqrt{\frac{T}{k}}\left(x^{+}_{P} - x_{Q}^{+}\right)\right]\cosh\left[\sqrt{\frac{\bar{T}}{k}}\left(x^{-}_{P} - x_{Q}^{-}\right)\right]
\nonumber\\
&
 - \left(\frac{ k}{\sqrt{T\bar{T}}}e^{\rho_{P}+\rho_{Q}} +  \frac{\sqrt{T\bar{T}}}{k}e^{-\left(\rho_{P}+\rho_{Q}\right)}\right)\sinh\left[\sqrt{\frac{T}{k}}\left(x^{+}_{P} - x_{Q}^{+}\right)\right]\sinh\left[\sqrt{\frac{\bar{T}}{k}}\left(x^{-}_{P} - x_{Q}^{-}\right)\right].
\end{align}

\noindent Bulk quantities in AdS/CFT are usually divergent as $\rho \to \infty$, reflecting the short distance (UV) divergences in the dual field theory. The simplest way to regulate such divergences is to place the boundary on a $\rho = \rho_{0}$ slice, with finite $\rho_0\,$, and take $\rho_0 \to \infty\,$ at the end. Let us then push the points $P$, $Q$ in \eqref{loop for AAdS3} to the regularized conformal boundary of AdS$_{3}$, i.e. $\rho_{P} = \rho_{Q} = \rho_{0}\to \infty\,$. Our basic observation is that the composite loop $W(P,Q)$ in the fundamental representation of $sl(2,\mathds{R})$ is related to the length $d(P,Q)$ of the geodesic anchored at $P$ and $Q$ as 
\begin{equation}\label{W and geodesic distance}
W_{2d}(P,Q) = 2 \cosh d(P,Q) \xrightarrow[\rho_{0}\to \infty]{} \exp d(P,Q)\,,
\end{equation}

\noindent where we used the fact that the geodesic length becomes large (divergent, in fact) as we push the points to the boundary. As we have discussed, for standard gravitational theories in the bulk the geodesic distance is intimately related to entanglement entropy in the dual theory via the R-T prescription. The functional \eqref{definition composite loop} rephrases this result in a language appropriate to the Chern-Simons theory, and it is moreover well-defined in the higher spin theory as well. Motivated by this fact, for points $P$ and $Q$ on a Cauchy slice on the boundary, defining a spacelike interval $A\,$, we propose to consider the functional
\begin{equation}\label{EE formula}
S_{A} \equiv \frac{k_{cs}}{\fudge}\log\left[\lim_{\rho_{0} \to \infty} W_{\mathcal{R}}(P,Q)\Bigr|_{\rho_{P}=\rho_{Q}=\rho_{0}}\right]
\end{equation}

\noindent as a candidate entanglement entropy in the $2d$ CFTs dual to the three-dimensional higher spin theories. Here, $k_{cs}$ is the Chern-Simons level defined in \eqref{kcs}. The constant $\fudge$ takes the value $2$ if there are half-integer spin currents in the spectrum, and $1$ otherwise; its origin will be explained in section \ref{non princip}. As we will discuss below, the choice of representation $\mathcal{R}$ in \eqref{EE formula} depends on both $N$ and the choice of embedding of $sl(2)$ into $sl(N)\,$. In the $N=2$ case, a change in the chosen representation can be compensated by changing the prefactor in \eqref{EE formula}. For example, evaluating \eqref{definition composite loop} in a three-dimensional representation of $sl(2)$, instead of the fundamental, one obtains $W_{3d}(P,Q) = -1 + \Bigl(W_{2d}(P,Q)\Bigr)^{2}  \xrightarrow[\rho_{0}\to \infty]{} \Bigl(W_{2d}(P,Q)\Bigr)^{2}$, and \eqref{EE formula} would remain invariant if we simultaneously divide the prefactor by two. Using the Chern-Simons level as the coefficient (up to $\fudge$) appears as a natural choice from the bulk perspective, which does not rely on details of the representation. Once the coefficient is fixed in this way, we will select the representation based on physical requirements. We emphasize that for $N>2$ the above functional does not have an obvious geometric interpretation: it is determined purely in terms of the gauge connections, as appropriate to the topological character of the bulk theory, and in particular it does not require the identification of a metric tensor.

A comment is in order about the symmetries that \eqref{EE formula} is expected to have. Under a general gauge transformation, a Wilson line $U(P,Q)=\mathcal{P}\exp\left(\int_{Q}^{P}A\right)$ transforms as $U \to h^{-1}(P)U(P,Q)h(Q)\,$. Since $A$ is valued in the $sl_L$ (``left") algebra and $\bar{A}$ is valued in the $sl_R$ (``right") algebra, it is clear that the composite loop $W_{\mathcal{R}}(P,Q)$ in \eqref{definition composite loop} is invariant under the diagonal subgroup parameterized by $h=\bar{h}\,$. As reviewed above, in the pure gravity case ($N=2$) the diagonal subgroup corresponds to rotations of the local Lorentz frame (c.f. \eqref{transformation vielbein}), so such invariance is very natural. Also, in the pure gravity case, off-diagonal gauge transformations correspond to a shift of the endpoints as is clear from (\ref{transformation vielbein}), and the geodesic length is obviously not invariant under such shifts. More importantly, even though  \eqref{definition composite loop} is not invariant under a general gauge transformation, in \eqref{EE formula} we are only assigning a field-theoretical interpretation to its leading UV (large $\rho_0$) divergence. Now recall that the asymptotic behavior of the gauge fields encodes the state of the system, and that gauge transformation that change the asymptotic behavior change the state of the system and are not true symmetries, whereas gauge transformations that leave the asymptotic behavior invariant are true symmetries. Gauge transformations of the first type belong to the so-called asymptotic symmetry group of the system. We therefore see that the entanglement entropy is only invariant under those gauge transformations that act trivially on the state, and not under those that modify the state, exactly as expected.

In the absence of explicit field theory results for entanglement entropy in 2$d$ CFTs deformed by higher spin currents (or in non-trivial states carrying higher spin charges), we will content ourselves with testing the plausibility of our proposal. Firstly, we will explicitly check that it allows us to recover the known CFT results in the absence of higher spin charges. Secondly, it will reproduce the thermal entropy in the limit in which the von Neumann entropy becomes extensive. Finally, we will check that the functional $S_{A}$ satisfies the strong subadditivity property of entanglement entropy (up to subtleties that we will discuss in due course).

%%%%%%%%%%%%%%%%%%%%%%%%%%%%%%%%%%%%%%%%%%%%%%%
\subsection{Recovering standard results}
%%%%%%%%%%%%%%%%%%%%%%%%%%%%%%%%%%%%%%%%%%%%%%%
We will now show that our prescription, when applied to solutions of pure gravity (i.e. in the absence of higher spin charges), allows us to recover the known results for the single-interval entanglement entropy in $2d$ CFTs \cite{Calabrese:2004eu}. With the result \eqref{loop for AAdS3} for AAdS$_{3}$ solutions in hand we can easily compute \eqref{EE formula} for the rotating BTZ black hole, characterized by \eqref{BTZ L and Lbar}, as well as global AdS$_{3}$ \eqref{AdS3 L and Lbar}, and the Poincar\'e-patch of AdS$_{3}$ (the latter with $T=\bar{T}=0$). From the dual field theory point of view,  the Poincar\'e-patch and global AdS$_{3}$ backgrounds correspond to the CFT ground state (i.e. at zero temperature) on the infinite line and on a system with periodic boundary conditions, respectively. The rotating black hole background, in turn, corresponds to computing the entanglement entropy in a finite temperature state with a potential for angular momentum. In the $N=2$ case \eqref{central charge} implies $c = 6k_{cs}$ and \eqref{EE formula} yields
\begin{align}
\mbox{Poincar\'e-patch:}&
&  S_{PAdS_{3}} &= \frac{c}{3}\log\left[\frac{\Delta x}{a}\right]
\\
 \mbox{global:}&
&S_{AdS_{3}} &= \frac{c}{3}\log\left[\frac{\ell}{a}\sin\left(\frac{\Delta \varphi}{2}\right)\right]
 \\
\mbox{black hole:}&
&S_{BTZ} &=
\frac{c}{6}\log\left[\frac{\beta_{+}\beta_{-}}{\pi^{2}a^{2}}\sinh\biggl(\pi \frac{\Delta x}{\beta_{+}}\biggr) \sinh\biggl(\pi\frac{\Delta x}{\beta_{-}}\biggr)\right]
\label{BTZ ent ent 2d}
\end{align}
\noindent where we defined the ``lattice spacing" $a$ in terms of the radial cutoff $\rho_{0}$ as
\begin{equation}
a = \ell e^{-\rho_{0}}\,,
\end{equation}

\noindent and dropped contributions that are subleading as $\rho_0 \to \infty\,$, as instructed by the limit in \eqref{EE formula}. In the results for Poincar\'e AdS and the BTZ black hole we have defined $\ell \varphi \to x\,$. Similarly, $\Delta x \equiv x_{P}-x_{Q}$ and $\Delta \varphi \equiv \varphi_{P}-\varphi_{Q}\,$. The zero temperature results, as well as the finite temperature result in the absence of rotation were first reproduced using holography in \cite{Ryu:2006bv}. The rotating case lies beyond the scope of the original R-T prescription, however, because in a stationary (but not static) spacetime the extremal surface anchored at the spacelike interval in the boundary does not necessarily lie on a constant-time slice in the bulk. The corresponding result was later obtained with the refined covariant prescription put forward in \cite{Hubeny:2007xt}. It is reassuring that our prescription in terms of Wilson lines encompasses all these cases simultaneously, and with a minimal calculational effort. 

A comment is in order regarding the black hole result \eqref{BTZ ent ent 2d}. As written, this result is valid for ``planar" black holes only. The same result holds for globally-defined black holes where the boundary is $S^1$, but only for sufficiently small $\Delta x$, see section~\ref{subsec:choice of rep}. It is somewhat remarkable that a universal answer exists even in this case, where the result is not determined by conformal symmetry. Since the black hole also has a temperature, the corresponding CFT is defined on a torus. Field-theoretical calculations of entanglement in $2d$ CFTs are usually performed using the so-called \textit{replica trick}, and (for the single-interval case) they effectively boil down to the calculation of a two-point function of twist operators (see e.g. \cite{Calabrese:2004eu}). In the cylinder (or the plane) the form of this two-point function is completely fixed by symmetry (Ward identities), and hence universal. As pointed out in \cite{Calabrese:2009qy}, however, the corresponding correlators on the torus depend not only on the conformal weights, but on specific details of the theory such as the operator content. From the holographic point of view, it is conceivable that these non-universalities are washed out in the semiclassical (large-$c$) limit, much in the same way that non-universalities associated with multiple intervals (rather than finite size effects) have been recently shown to be subleading in the large central charge regime \cite{Hartman:2013mia,Faulkner:2013yia}.

We stress that even though our formula \eqref{EE formula} correctly reproduces the entanglement entropy in the absence of higher spin charges and chemical potentials, there is no \textit{a priori} guarantee that it will still compute the entanglement entropy in the higher spin cases. In what follows we will amass more evidence in favor of this interpretation.

%%%%%%%%%%%%%%%%%%%%%%%%%%%%%%%%%%%%%%%%%%%%%%%
\subsection{Thermal entropy and the choice of representation}\label{subsec:choice of rep}
%%%%%%%%%%%%%%%%%%%%%%%%%%%%%%%%%%%%%%%%%%%%%%%

When computed at finite temperature, the von Neumann entropy associated with the reduced density matrix $\rho_{A}$ receives contributions from classical correlations that mix with the quantum correlations due to ``true" entanglement. For fixed temperature, as subsystem $A$ grows in size the reduced density matrix approaches the thermal density matrix of the full system; by definition, the entanglement entropy then becomes the thermal entropy (up to subtraction of short distance divergences).  In our discussion of this limit we will distinguish between a system that is infinitely-extended in the spatial direction, and one that is compact (i.e. a circle). In the holographic context the former arises as the boundary of a planar black hole, while the latter corresponds to a globally-defined black hole geometry. 

\begin{figure}[h]
\centering
\includegraphics[width=15cm]{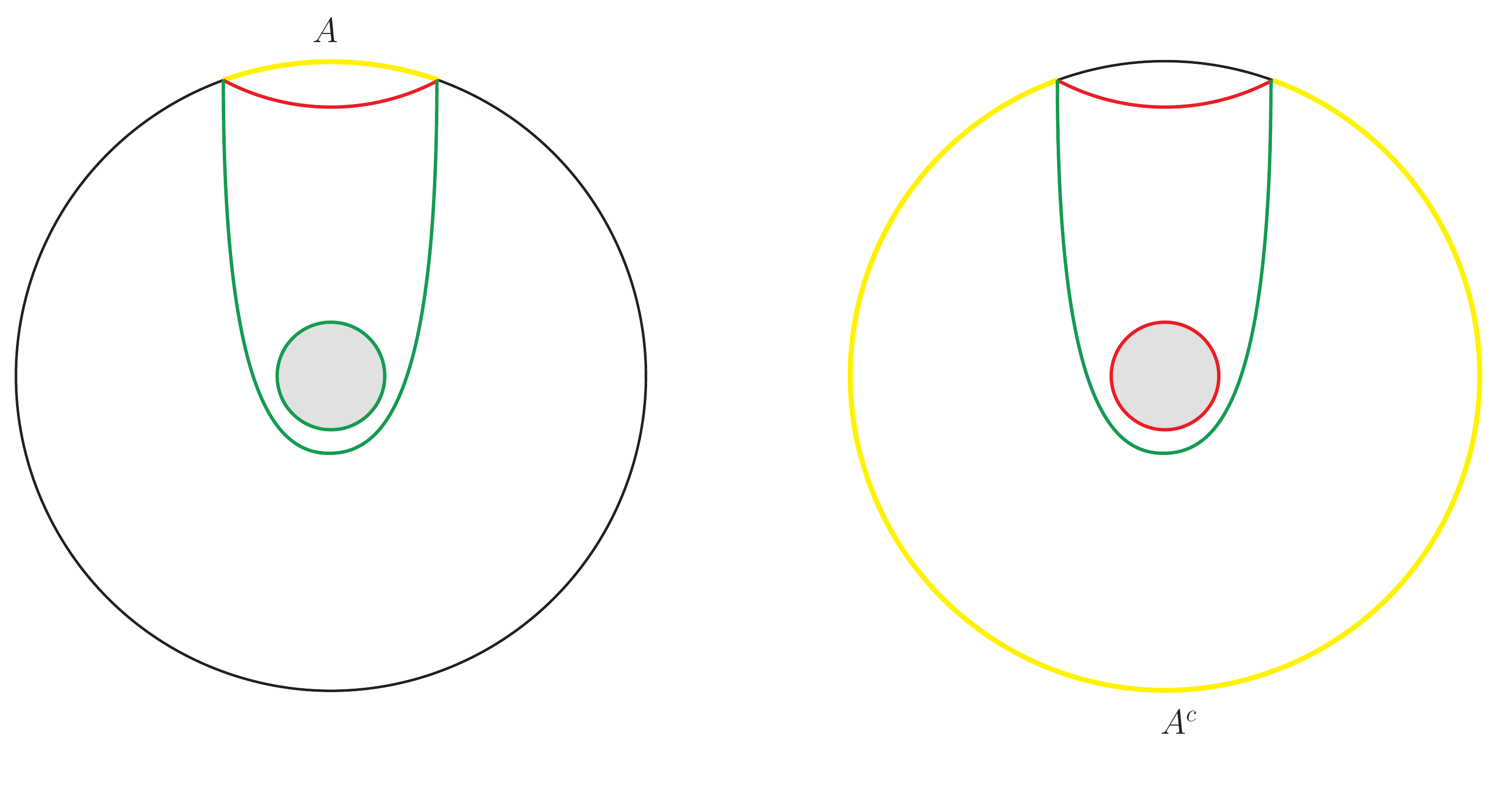}
\caption{Minimal surface with non-trivial bulk topology in the R-T prescription. The interior of the black hole horizon is represented by the grey shaded area. Left figure: for a small boundary region (yellow), the minimal surface (red) is given by a connected geodesic. Right figure: for a large boundary region (yellow), the minimal surface (red) is disconnected and includes a component that wraps around the horizon, effectively computing its area and hence the black hole thermal entropy.}
\label{fig:Compact}
\end{figure}

Let us start with the case of a compact system. The cycle parameterizing the horizon of a global black hole is non-contractible, and the non-trivial topology of the bulk manifold in this case makes the definition of holographic entanglement entropy subtle. Let us first review this issue in the context of the R-T proposal, and the global BTZ black hole. As depicted in figure \ref{fig:Compact}, in the presence of a black hole there are in general two geodesic configurations that are homologous to an interval in the boundary. For a fixed temperature (i.e. fixed horizon size), if the boundary interval is small, the corresponding minimal surface is a connected geodesic that does not wrap the horizon. For a sufficiently large boundary interval, the minimal surface will instead be a disconnected sum of two components, one of which is a loop around the black hole horizon \cite{Azeyanagi:2007bj}. The length of the latter curve effectively computes the black hole horizon area (length), and hence its thermal entropy via the Bekenstein-Hawing formula. Equivalently, for a fixed interval size, the minimal surface can change from connected to disconnected as a function of the size of the horizon (temperature). Incidentally, this shows that the limits of high temperature and large subsystem size do not commute.\footnote{We thank Mukund Rangamani for emphasizing the importance of this issue to us.} 

We now ask how is the thermal entropy recovered in the context of our proposal, focusing on the global BTZ solution (c.f. \eqref{BTZ L and Lbar}) as an example. The entropy of the BTZ black hole (and of the dual theory defined on a torus) is given by
 \begin{align}
S_{\textit{\scriptsize{thermal}}}  =
 2\pi\sqrt{k \stress} + 2\pi \sqrt{ k \stressbar } = 2\pi\sqrt{\frac{c}{6} \stress} + 2\pi \sqrt{ \frac{c}{6}  \stressbar } \,,
\end{align}
\noindent which moreover takes the form predicted by Cardy's asymptotic growth of states in a unitary CFT. Starting with the result \eqref{loop for AAdS3}, we evaluate it for an equal-time loop in the $\varphi$ direction, at a generic radial position $\rho\,$, and extremize the result as a function of $\rho\,$. The value $\rho_*$ that minimizes this functional is such that
\begin{equation}
e^{2\rho_*} = \frac{\pi^{2}\ell^{2}}{\beta_+\beta_+}\,,
\end{equation}

\noindent which in fact corresponds to the (outer) black hole horizon $\rho_+$. Evaluating \eqref{loop for AAdS3} at $\rho_* = \rho_{+}$ we find
\begin{equation}
W(\Delta \varphi = 2\pi)\Bigr|_{\rho = \rho_*} = 2\cosh\left[\frac{S_{\textit{\scriptsize{thermal}}} }{k}\right],
\end{equation}

\noindent so that
\begin{equation}
S_{\textit{\scriptsize{thermal}}}  = k \cosh^{-1}\left(\frac{1}{2}W(\Delta \varphi = 2\pi)\Bigr|_{\rho = \rho_*}\right).
\end{equation}

\noindent Notice that the appearance of the $\cosh^{-1}$ function is consistent with the first equality in \eqref{W and geodesic distance}, and the fact that we are evaluating at a finite radial distance instead of pushing the points to the boundary.

While it is not obvious how to generalize the above minimization procedure to the higher spin scenario,\footnote{In particular, it is not clear what should replace the inverse cosh, if anything. However, since we expect closed loops that wrap around the horizon to reproduce the thermal entropy of the black hole, and we have an explicit expression for the latter, a working hypothesis could be to assign the black hole entropy to all closed bulk loops and to use our Wilson loop prescription for all curves that start and end on the boundary.} we will now show that our prescription can recover the thermal entropy in situations where the boundary coordinate is non-compact, even in the presence of higher spin charges. Let us first recall the corresponding result in field theory. Consider a $(1+1)$-dimensional CFT on the infinite line, in an ensemble at temperature  $\beta^{-1}\,$, and let region $A$ be an interval of length $\Delta x$: the basic idea is that, up to a proper subtraction of ultraviolet divergences, the entanglement entropy should coincide with the thermal entropy in the limit $\Delta x \gg \beta\,$. More explicitly, consider the finite temperature result for the single-interval entanglement entropy,
\begin{equation}\label{finite temp ent ent}
S_{A} = \frac{c}{3}\log\left(\frac{\beta}{\pi a}\sinh\left(\frac{\pi \Delta x}{\beta}\right)\right).
\end{equation}
\noindent In the limit $\Delta x \gg \beta$  the entanglement entropy $S_{A}$ becomes extensive
\begin{equation}\label{extensive thermal entropy}
S_{A} \simeq \frac{\pi c}{3}\frac{\Delta x}{\beta} \,,
\end{equation}

\noindent with a coefficient given by the thermal entropy density of the system.

 In standard gravity, the thermal entropy associated with a black hole is computed from the area of the black hole horizon using the Bekenstein-Hawking formula $S_{\textit{\scriptsize{thermal}}} = \mbox{Area}/(4G)$. The notion of a smooth horizon is not invariantly defined in the topological bulk theory, and hence the entropy must be computed by different means when the gauge group is $SL(N,\mathds{R})\times SL(N,\mathds{R})$ with $N>2\,$. In \cite{deBoer:2013gz} we showed that, for any $N$, the higher spin black hole thermal entropy can be written in terms of the connection as
\begin{equation}\label{our entropy formula}
S_{\textit{\scriptsize{thermal}}} = -2\pi i k_{cs}\,\text{Tr}_{N}\Bigl[\left(a_{z}+a_{\bar{z}}\right)\left(\tau a_{z} +\bar{\tau}a_{\bar{z}}\right)-\left(\bar{a}_{z}+\bar{a}_{\bar{z}}\right)\left(\tau\bar{a}_{z} + \bar{\tau}\bar{a}_{\bar{z}}\right)\Bigr],
\end{equation}

\noindent where the trace is taken in the fundamental representation. This is obtained by evaluating the free energy with canonical boundary conditions, and performing a Legendre transform (see \cite{Banados:2012ue} also). There also exists a different expression for the entropy which we called the holomorphic entropy 
formula in \cite{deBoer:2013gz} and which appears to be more closely connected to CFT partition functions. Our expression for the entanglement entropy favors the canonical version of the entropy over the holomorphic one, and we will return to this point in the conclusions.

In \cite{Gutperle:2011kf,Ammon:2011nk} it was proposed that a gauge-invariant characterization of a smooth black hole solution is the requirement that it has trivial holonomy around the contractible cycle of the boundary torus.\footnote{More precisely, by ``trivial" we mean that the holonomy is contained in the center of the gauge group \cite{Castro:2011fm}.}  In particular this means that the holonomy matches that of the BTZ black hole, and this requirement translates into $\left(\tau a_{z} +\bar{\tau}a_{\bar{z}}\right)=u^{-1} \left(i \Lambda^{0}\right)u\,$ for some matrix $u$. Since $a_{x} = \left(a_{z}+a_{\bar{z}}\right)/\ell$ and $\left(\tau a_{z} +\bar{\tau}a_{\bar{z}}\right)$ commute by the equations of motion (for constant connections), they can be diagonalized simultaneously, and the entropy density $s_{\textit{\scriptsize{thermal}}}$ reduces to
\begin{equation}\label{our Cardy formula}
s_{\textit{\scriptsize{thermal}}}  \equiv \frac{S_{\textit{\scriptsize{thermal}}}}{2\pi \ell} =k_{cs}\text{Tr}_{N}\Bigl[\left( \lambda_{x} - \overline{\lambda}_{x}\right)\Lambda^{0}\Bigr],
\end{equation}

\noindent where $\lambda_{x}$ and $\overline{\lambda}_{x}$ are diagonal matrices whose entries contain suitably ordered
eigenvalues of $a_{x}$ and $\bar{a}_{x}\,$, respectively. In what follows we will argue that, provided the representation $\mathcal{R}$ in \eqref{EE formula} is chosen appropriately, our result for the single-interval entanglement entropy will satisfy
\begin{equation}
S_{A} \xrightarrow[\Delta x \,\gg\, \beta]{} s_{\textit{\scriptsize{thermal}}}\, \Delta x
\end{equation}

\noindent in the extensive limit $\Delta x \,\gg\, \beta\,$. We will divide the discussion into principal and non-principal embeddings.

\subsubsection{Principal embedding}
In the principal embedding, $\Lambda^0$ is a diagonal matrix whose entries correspond to the components of the Weyl vector $\vec{\rho}$ of $sl(N)\,$ (c.f. appendix \ref{sec:reptheory}). Similarly, $a_{x}$ and $\bar{a}_{x}$ can be put in the Cartan subalgebra $\mathcal{C}$ by conjugation with a group element, and we denote the corresponding dual element in $\mathcal{C}^{*}$ (the root space) by $\vec{\lambda}_{x}$, $\vec{\overline{\lambda}}_{x}$. Hence, we can rewrite the entropy density \eqref{our Cardy formula} more abstractly as
\begin{equation}\label{our Cardy formula v2}
s_{\textit{\scriptsize{thermal}}}  =  k_{cs}\langle \vec{\lambda}_{x}-\vec{\overline{\lambda}}_{x}\,,\vec{\rho}\, \rangle\,.
\end{equation}

\noindent  On the other hand, in a given representation $\mathcal{R}$, the product of exponentials in \eqref{definition composite loop} contains a sum over terms of the form $e^{\Delta x \langle \vec{\lambda}_{x} - \vec{\overline{\lambda}}_{x},\lambda_{\mathcal{R}}^{(j)}\rangle}\,$, where $\lambda_{\mathcal{R}}^{(j)}$ denotes the weights in the corresponding representation. The question is now which of these terms dominates for large $\Delta x$. The vector $\vec{\lambda}_{x} - \vec{\overline{\lambda}}_{x}$ can, possibly up to a Weyl reflection, always be written as a sum of fundamental weights with non-negative coefficients. This is quite easy to see if we visualize $\vec{\lambda}_{x} - \vec{\overline{\lambda}}_{x}$ as a diagonal $N\times N$ matrix. The Weyl group permutes the diagonal entries arbitrarily, and in particular there always exists a permutation that orders the diagonal entries from larger to smaller. Such matrices precisely correspond to sums of fundamental weights with non-negative coefficients. If $\vec{\lambda}_{x} - \vec{\overline{\lambda}}_{x}$ is of this form, then the highest weight of the representation ${\cal R}$ will dominate the entanglement entropy for large $\Delta x$. All other weights are related to the highest weight by subtracting a combination of positive roots, and this will always lower the inner product. Therefore, up to a possible Weyl reflection, \eqref{EE formula} becomes
\begin{equation}
S_{A} \xrightarrow[\Delta x \, \gg \, \beta]{}
  k_{cs} \langle \vec{\lambda}_{x} - \vec{\overline{\lambda}}_{x}\,,\Lambda^{hw}_{\mathcal{R}}\rangle\,\Delta x\,,
\end{equation}
\noindent where $\Lambda^{hw}_{\mathcal{R}}$ denotes the highest weight in the representation $\mathcal{R}$. Comparing with \eqref{our Cardy formula v2},
and keeping in mind that close to the BTZ point (where the higher spin charges and chemical potentials vanish) $\vec{\lambda}_{x} - \vec{\overline{\lambda}}_{x}$ is a small perturbation of a multiple of $\Lambda^0$ and therefore automatically of the right form without the need for a Weyl reflection, 
we conclude that, in the principal embedding, our entanglement functional will correctly reproduce the thermal entropy in the Cardy limit provided we evaluate $W(P,Q)$ in the representation with highest weight given by the Weyl vector, i.e.
\begin{equation}\label{prin embedd rep}
\text{principal embedding:}\quad \Lambda^{hw}_{\mathcal{R}} = \vec{\rho}\,.
\end{equation}

\noindent Via the Weyl formula, the dimension of this representation is 
\begin{equation}
\text{dim}(\mathcal{R})=\prod_{\alpha >0}\frac{\langle \Lambda_{\mathcal{R}}^{hw} + \vec{\rho} ,\alpha\rangle}{\langle \vec{\rho},\alpha \rangle} = 2^{\# \text{of positive roots}} = 2^{\frac{N(N-1)}{2}}\,.
\end{equation}

\noindent Naturally, for $N=2$ this is the two-dimensional (defining) representation. For $N=3$ we have $\text{dim}(\mathcal{R}) = 8\,$; hence, in the $sl(3,\mathds{R})\times sl(3,\mathds{R})$ theory with principally-embedded $sl(2)$, one should evaluate \eqref{EE formula} in the adjoint representation in order to recover the thermal entropy in the high-temperature limit, and we will explicitly check this below by applying our formula to the spin-3 black hole. 

In addition to giving the right thermodynamic entropy, it is not hard to see that the representation \eqref{prin embedd rep} is the one needed for \eqref{EE formula} to yield the right result when applied to the principally-embedded BTZ solution in the higher spin theory. Ultimately this can be traced back to the factor $\text{Tr}_{N}\left[\Lambda^0\Lambda^0\right]$ in \eqref{central charge}, which in the principal embedding evaluates to the square of the Weyl vector. 

\subsubsection{Non-principal embeddings}\label{non princip}

The combined requirements that the entanglement entropy reproduces the thermal entropy at high temperature and that the BTZ result is recovered with the right normalization allow one to determine the representation $\mathcal{R}$ in non-principal embeddings as well. Let $\lambda_0$ denote the dual of the Cartan element $\Lambda^0\,$. If the embedding is such that the spectrum contain half-integer spin currents, $\lambda_0$ does not belong to the weight lattice, but $2\lambda_0$ does. The factor $\fudge $ introduced in \eqref{EE formula} accounts for this fact: for any embedding, $\fudge \lambda_0$ is a combination of fundamental weights with integer coefficients, so it proves convenient to define
\begin{equation}
\tilde{\lambda}_{0}=\fudge \lambda_0\,.
\end{equation}
\noindent The basic observation that leads to the choice of representation is that, at the BTZ point, $\lambda_{x} \sim \Lambda^0\,$. However,
this is not yet of the form of a sum of fundamental weights with non-negative coefficients and we still need to find a Weyl reflection that
puts it in this form. Let us denote this Weyl reflection by $w$. Then
in order to reproduce the thermal entropy \eqref{our Cardy formula}, we want the overlap between $w(\tilde{\lambda}_{0})$ and the highest weight appearing in the representation $\mathcal{R}$ to be the same as the overlap of $\tilde{\lambda}_0$ with itself,
\begin{equation}\label{overlap condition}
\langle \Lambda^{hw}_{\mathcal{R}}, w(\tilde{\lambda}_0)\rangle = \langle \tilde{\lambda}_0, \tilde{\lambda}_0\rangle =\fudge^2\text{Tr}_{N}\left[\Lambda_{(P)}^0\Lambda_{(P)}^0\right],
\end{equation}

\noindent where $\text{Tr}_{N}$ denotes the trace evaluated in the defining ($N$-dimensional) representation, and $\Lambda^0_{(P)}$ is the $\Lambda^0$ generator appropriate to the embedding labeled by the partition $P\,$. Similarly, in order to have an unambiguous thermal limit, we require that all the other states occurring in the representation have a strictly smaller overlap with $\tilde{\lambda}_0\,$, which will be automatically the case for the correct choice
of Weyl reflection $w$. As we deform the theory away from the BTZ point by adiabatically turning on the higher spin charges along the BTZ branch, and as long as we do not encounter eigenvalue crossing, these requirements will still select the right representation.

Given an embedding $P$, we therefore find that the unique representation satisfying the above requirements is the one whose highest weight state is given by the (unique) dominant weight\footnote{A dominant weight is a linear combination of fundamental weights with non-negative coefficients.} that lies in the same Weyl orbit as $\tilde{\lambda}_0\,$. For the first few values of $N$, the representations selected by this criteria are shown in table \ref{table reps}. In particular, for any $N$, the desired representation in the diagonal embedding is the adjoint, with highest weight $\Lambda^{hw}_{\mathcal{R}} = \alpha_{1} =\omega_{1} + \omega_{N-1}\,$.

\begin{table}[h]
\begin{center}
\begin{tabular}{|l|c|c|c|}
\hline 
\multicolumn{1}{|c|}{Embedding} & $\fudge$ & Highest weight & $\text{dim}(\mathcal{R})$ \\ 
\hline 
$\mathbf{3} \to \mathbf{2}_{2} \oplus \mathbf{1}_{2}$ & $2$ &$ \omega_{1}+\omega_{2}$ & 8 \\ 
\hline 
$\mathbf{4} \to \mathbf{2}_{2} \oplus \mathbf{1}_{2}\oplus \mathbf{1}_{2}$ &  $2$ & $ \omega_{1}+\omega_{3}$ & 15 \\ 
\hline 
$\mathbf{4} \to \mathbf{2}_{2} \oplus \mathbf{2}_{2}$ & 1 &$ \omega_{2}$ & 6 \\ 
\hline 
$\mathbf{4} \to \mathbf{3}_{2} \oplus \mathbf{1}_{2}$ & 1 &$ \omega_{1}+\omega_{3}$ & 15 \\ 
\hline 
$\mathbf{5} \to \mathbf{2}_{2} \oplus \mathbf{1}_{2} \oplus \mathbf{1}_{2} \oplus \mathbf{1}_{2}$ &   $2$ &$ \omega_{1}+\omega_{4}$ & 24 \\ 
\hline
$\mathbf{5} \to \mathbf{2}_{2} \oplus \mathbf{2}_{2} \oplus \mathbf{1}_{2}$ & $2$ &$ \omega_{2}+\omega_{3}$ & 75 \\ 
\hline
$\mathbf{5} \to \mathbf{3}_{2} \oplus \mathbf{1}_{2} \oplus \mathbf{1}_{2}$ &1 &$ \omega_{1}+\omega_{4}$ & 24 \\ 
\hline 
$\mathbf{5} \to \mathbf{3}_{2} \oplus \mathbf{2}_{2}$ &  $2$ &$ \omega_{1}+\omega_{2}+\omega_{3}+\omega_{4}$ & 1024 \\ 
\hline 
$\mathbf{5} \to \mathbf{4}_{2} \oplus \mathbf{1}_{2}$ &   $2$  &$ 2\omega_{1}+\omega_{2} +\omega_{3} + 2\omega_{4}$ & 6125 \\ 
\hline 
\end{tabular}  
\end{center}
\caption{Choice of representation in the non-principal embeddings for $N=3,4,5\,$.}
\label{table reps}
\end{table}

%%%%%%%%%%%%%%%%%%%%%%%%%%%%%%%%%%%%%%%%%%%%%%%
\subsection{Strong subadditivity} \label{subsec: strong sub}
%%%%%%%%%%%%%%%%%%%%%%%%%%%%%%%%%%%%%%%%%%%%%%%
An important property of entanglement entropy is the so-called \textit{strong subadditivity} \cite{lieb:1938}:
\begin{align}\label{strong sub}
S_A + S_B \geq{}&
 S_{A\cup B} + S_{A \cap B}
\\
S_{A} + S_{B} \geq{}&
 S_{A \cap B^{c}} + S_{B \cap A^{c}}
 \label{strong sub 2}
\end{align}
\noindent One of the main successes of the R-T prescription is its ability to fulfill these inequalities in a natural way \cite{Hirata:2006jx,Headrick:2007km,Wall:2012uf}. In order to discuss whether the functional \eqref{EE formula} satisfies strong subadditivity we must distinguish two cases: $A \cap B = \emptyset$ (i.e. $A$ and $B$ are disjoint intervals) and $A \cap B \neq \emptyset\,$. We will only study situations where the topology of the bulk is trivial, i.e. there are no non-contractible cycles such as those associated with global black holes.

\subsubsection{Disjoint intervals}
\begin{figure}[h]
\centering
\includegraphics[width=15.7cm]{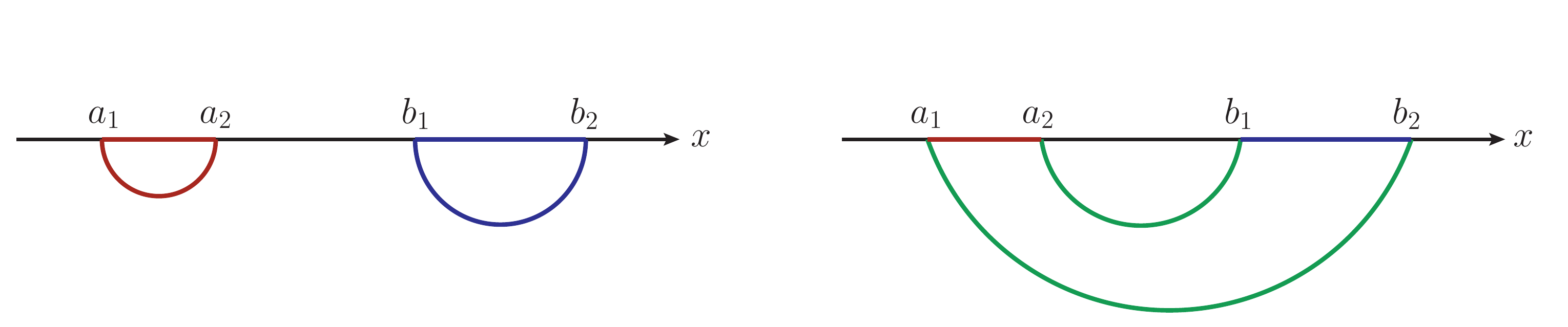}
\caption{Relevant configurations for two disjoint intervals on the boundary. The pairing $(a_{1},b_{1})\,$ and $(a_{2},b_{2})$ is excluded by a condition on the homology of the bulk configuration.}
\label{fig:SSA1}
\end{figure}

Firstly, we need to supplement our prescription with an instruction on how to compute the entanglement for a region of the form $A\cup B$ when $A$ and $B$ are disjoint intervals. Inspired by the pure gravity case, we propose to minimize the result over all the possible pairings between the boundary points defining the intervals $A$ and $B$, and such that the bulk configuration is homologous to the boundary. The latter topological condition was originally introduced in \cite{Fursaev:2006ih}, and it has been shown to be a necessary ingredient for the consistency of the Ryu-Takayanagi prescription \cite{Headrick:2007km}. In the present context, it implies that the topology of the configuration for which \eqref{definition composite loop} is evaluated is such that there exists a bulk region bounded by the Wilson lines and the boundary intervals $A$ and $B\,$. For example, if $A$ and $B$ are two disjoint intervals defined by boundary points $(a_{1},a_{2})$ and $(b_{1},b_{2})\,$, the two relevant configurations are depicted in figure \ref{fig:SSA1}. We then \textit{define} the quantity $S_{A \cup B}$ as
\begin{equation}\label{min instruction}
S_{A \cup B} = \min \Bigl\{S(a_{1},a_{2})+S(b_{1},b_{2})\,,\,S(a_{1},b_{2})+S(a_{2},b_{1})\Bigr\},
\end{equation}

\noindent where the individual terms in each sum are computed using \eqref{EE formula}. Note that the pairing that would give $S(a_{1},b_{1})+S(a_{2},b_{2})$ is excluded by the homology condition. With the definition \eqref{min instruction}, \eqref{strong sub} is automatically satisfied when $A \cap B = \emptyset \,$ (\eqref{strong sub 2} is satisfied trivially in this case).

\noindent Naturally, these considerations can be generalized to any number of disjoint intervals.

\subsubsection{Overlapping intervals} \label{overlapping strong sub}
\begin{figure}[h]
\centering
\includegraphics[width=15cm]{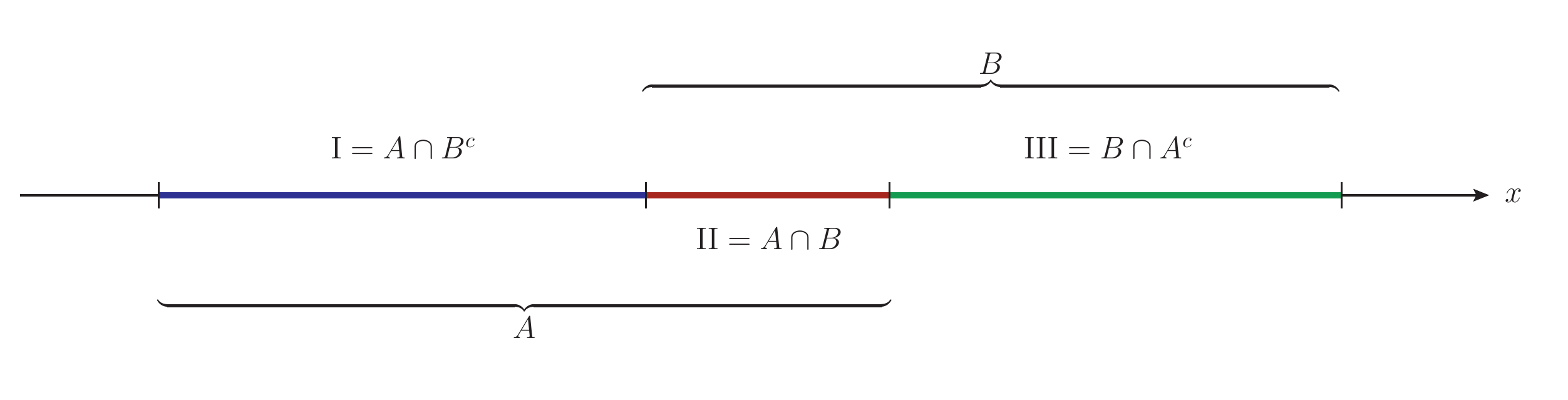}
\caption{Configuration for overlapping intervals.}
\label{fig:SSA2}
\end{figure}

Let us now consider \eqref{strong sub}-\eqref{strong sub 2} in the case where the regions $A$ and $B$ intersect. As we will now show, these inequalities are satisfied if the single-interval entanglement entropy is a concave and non-decreasing function of the interval length. The relevant boundary configuration in this case is depicted in figure \ref{fig:SSA2}. Let $\Delta_{\text{I}}$, $\Delta _{\text{II}}$ and $\Delta_{\text{III}}$ denote the length of the corresponding intervals in figure \ref{fig:SSA2}. Following \cite{Callan:2012ip} we define
\begin{equation}
\lambda = \frac{\Delta _{\text{III}}}{\Delta _{\text{I}}+\Delta _{\text{III}}}\,,
\end{equation}

\noindent so that 
\begin{equation}\label{overlapping lenghts}
\Delta _{\text{I}}+\Delta _{\text{II}} = \lambda \Delta_{\text{II}} + (1-\lambda) \left(\Delta_{\text{I}}+\Delta _{\text{II}}+\Delta _{\text{II}}\right)\,,\qquad \Delta _{\text{II}}+\Delta _{\text{III}} = \lambda \left(\Delta_{\text{I}}+\Delta _{\text{II}}+\Delta _{\text{II}}\right) + (1-\lambda)\Delta_{\text{II}}\,.
\end{equation}

\noindent Assuming that the single-interval entanglement entropy is a concave function of the interval length, \eqref{overlapping lenghts} implies
\begin{align}
S_{A} \geq{}&
 \lambda S_{A\cap B} + (1-\lambda) S_{A \cup B} \,,
\\
 S_{B} \geq{}&
  \lambda S_{A\cup B} + (1-\lambda)S_{A\cap B}\,,
\end{align}

\noindent and adding these two inequalities we obtain \eqref{strong sub}. Similarly,  under the assumption that the single-interval entanglement is a non-decreasing function of the interval length one has
\begin{align}
S_{A} \geq
 S_{A\cap B^{c}} \,\,,\qquad 
 S_{B } \geq
  S_{B\cap A^{c}}\,,
\end{align}

\noindent and adding these two inequalities yields \eqref{strong sub 2}. 

Given that a general proof of the concavity and monotonicity of the functional \eqref{EE formula} has eluded us so far, in section \ref{sec: sl3 examples} we will study whether these properties are fulfilled on a case-by-case basis when applying \eqref{EE formula} to higher spin examples.

%%%%%%%%%%%%%%%%%%%%%%%%%%%%%%%%%%%%%%%%%%%%%%%%%%%%%%%%%%%%%%%%%%%%%%%%%%%%%%%%
\section{Examples in the  $SL(3,\mathds{R})\times SL(3,\mathds{R})$ theory}\label{sec: sl3 examples}
%%%%%%%%%%%%%%%%%%%%%%%%%%%%%%%%%%%%%%%%%%%%%%%%%%%%%%%%%%%%%%%%%%%%%%%%%%%%%%%%
Having checked that our entanglement functional reproduces all the known results for the standard gravity case, corresponding to $SL(2,\mathds{R})\times SL(2,\mathds{R})$ gauge group, we will now evaluate it on different solutions of the $SL(3,\mathds{R})\times SL(3,\mathds{R})$ theory. All the solutions we consider below are of the form \eqref{general form connections}, with $a$, $\bar{a}$ constant connections, which includes black hole solutions carrying higher spin charges. From now on we work in units in which the AdS length is set to one, $\ell =1\,$.

%%%%%%%%%%%%%%%%%%%%%%%%%%%%%%%%%%%%%%%%%%%%%%%
\subsection{The RG flow solution}\label{RG flow}
%%%%%%%%%%%%%%%%%%%%%%%%%%%%%%%%%%%%%%%%%%%%%%%
As a first non-trivial example we apply our formula to compute the entanglement entropy for a zero temperature solution discussed in \cite{Ammon:2011nk}, which realizes a flow from a UV CFT with $\mathcal{W}_{3}^{(2)}$ symmetry to an IR fixed point  with $\mathcal{W}_{3}$ symmetry. The flow is initiated by adding a relevant operator of dimension $3/2$ to the Lagrangian of the UV CFT. From the point of view of the theory with $\mathcal{W}_{3}$ symmetry, on the other hand, the flow is triggered by adding weight-3 currents to the Lagrangian of the IR CFT. 

As reviewed in section \ref{sec:CS gravity}, there are two non-trivial embeddings of the gravitational sector in the $SL(3,\mathds{R})\times SL(3,\mathds{R})$ theory. The corresponding theories have different spectrum and asymptotic symmetries; from the bulk perspective, they are constructed as excitations around different AdS vacua. In particular, the theory constructed around the principal embedding vacuum contains irrelevant operators, and it is interesting to ask whether it is possible to realize an RG flow from the theory in the diagonal embedding (UV), with central charge $c_{UV} = c/4$, to the principal embedding fixed point (IR) with central charge $c_{IR}=c\,$. This was accomplished in \cite{Ammon:2011nk}, where it was pointed out that even though $c_{IR} > c_{UV}$, such flow is triggered by adding operators which are not Lorentz invariant, and therefore there is no a priori contradiction with the $c$-theorem whose derivation assumes a Lorentz-invariant flow. From the point of view of the Chern-Simons theory the flow is realized by constructing a connection that interpolates between those corresponding to the UV and IR AdS$_{3}$ vacua. Using the basis of generators $\{L_{i},W_{j}\}$ introduced in appendix \ref{sec:algebras}, the RG flow solution reads
\begin{equation}
\begin{aligned}\label{RG flow connections 1}
A 
={}&
 e^{\rho}\,\hat{\Lambda}^{+}\,dx^{+} + \hat{\Lambda}^{0}\,d\rho +  \lambda\,e^{\rho/2} L_{1} \,dx^{-} 
\\
\bar{A} ={}&
 - e^{\rho}\,\hat{\Lambda}^{-}\, dx^{-} - \hat{\Lambda}^{0}\,d\rho   -\lambda\, e^{\rho/2} L_{-1}\,dx^{+}\,,
%  \label{RG flow connections 2}
\end{aligned}
\end{equation}

\noindent where
\begin{equation}\label{diag embed sl2 generators}
\hat{\Lambda}^{0} = \frac{1}{2}L_{0}\,,\qquad \hat{\Lambda}^{\pm} = \pm \frac{1}{4}W_{\pm 2}
\end{equation}

\noindent is the basis of $sl(2,\mathds{R})$ generators appropriate to the diagonal embedding (UV theory). More precisely, we have rewritten the solution in a way that looks natural from the point of view of the diagonal embedding. Namely, for $\lambda = 0$ the above solution is the UV AdS$_{3}$ vacuum, while for $\lambda \to \infty$ it gives the IR vacuum only after rescaling $\rho$ and exchanging $x^+ \leftrightarrow x^-\,$. As discussed in \cite{Ammon:2011nk}, from the point of view of the UV CFT with $\mathcal{W}_{3}^{(2)}$ symmetry $\lambda$ is interpreted as a source for spin-$3/2$ operators. We stress that the deep IR corresponds to large $\lambda$, and the UV to small $\lambda$.

As we explained in section \ref{subsec:choice of rep}, for the $N=3$ theory in the diagonal embedding we must evaluate \eqref{definition composite loop} and \eqref{EE formula} in the eight-dimensional adjoint representation (and with $\fudge =2$). For two arbitrary boundary points $(P,Q)$ we obtain
\begin{align}\label{RG flow loop}
W_{Adj}(P,Q) ={}&
 8 -24\lambda^2 e^{\rho_0} \Delta x^+ \Delta x^- - 2e^{2\rho_0}\Delta x^+\Delta x^-\Bigl(3-11\lambda^4 \Delta x^+\Delta x^-\Bigr)
\nonumber\\
&
+8\lambda^2 e^{3\rho_0} \left( \Delta x^+ \Delta x^-\right)^2\Bigl(1-\lambda^4  \Delta x^+ \Delta x^-\Bigr) 
\nonumber\\
&
+ e^{4\rho_0}\biggl[\lambda^8  \left(\Delta x^+ \Delta x^-\right)^4 - \lambda^4\left(\left(\Delta x^+ \right)^6+\left( \Delta x^-\right)^6\right)+\left(\Delta x^+ \Delta x^-\right)^2\biggr]
\end{align}
\noindent where $\Delta x^\pm = x_{P}^\pm - x_{Q}^{\pm}$ and  $\rho_0 \to \infty$ is the position of the regularized conformal boundary as before. Note that the above expression depends only on the invariant interval $\Delta s^{2} =-\Delta x^+\Delta x^-$ for $\lambda=0$ and $\lambda=\infty$, consistent with Lorentz invariance at the fixed points. The relativistic invariance is broken for generic values of $\lambda\,$. Evaluating for points $P$ and $Q$ at equal times we obtain
\begin{align}\label{RG flow loop 2}
W_{Adj}(P,Q) = 8 + \left(\frac{\Delta x}{e^{-\rho_0}}\right)^2\Biggl[&
e^{2\rho_0}  \left(\Delta x\right)^2\left(1 - \lambda^{4} \left(\Delta x\right)^2\right)^2+8\lambda^2 e^{\rho_0} \left(\Delta x\right)^2\left(1 + \lambda^4  \left(\Delta x\right)^2\right)
\nonumber\\
&+2\left(3+11\lambda^4 \left(\Delta x\right)^2\right)+24\lambda^2 e^{-\rho_0}  \Biggr]
\end{align}

\noindent where $\Delta x$ is the spatial separation between the points (i.e. the interval length). Focusing on the leading $\rho_0$-divergence, \eqref{EE formula} yields
\begin{equation}\label{EE for RG flow}
S_{A} = \frac{c_{UV}}{3}\log \left[\frac{\Delta x}{a_{UV}}\sqrt{\,\left| 1 -\lambda^{4}\left(\Delta x\right)^2\right| }\,\right]
\end{equation} 

\noindent where the central charge is given by \eqref{central charge},
\begin{equation}
c_{UV} = 12k_{cs}\text{Tr}_{3d}\left[\hat{\Lambda}^{0}\hat{\Lambda}^{0}\right] = 6k_{cs}\,,
\end{equation}

\noindent and the UV cutoff is defined as $a_{UV} \equiv e^{-\rho_0}\,$.  We notice that the appropriate result $S_{A} = (c_{UV}/3)\log[\Delta x/a_{UV}]$ is recovered at the UV fixed point $\lambda = 0\,$. As we increase $\Delta x$ from its lower bound $\Delta x = a_{UV}$ the derivative of  \eqref{EE for RG flow} is discontinuous at the value $\Delta x = \frac{1}{\sqrt{2}\lambda^{2}}$ and the putative entanglement entropy ceases to be non-decreasing at that point, which would conflict with strong subadditivity (c.f. section \ref{overlapping strong sub}). This discontinuity as a function of interval length indicates that \eqref{EE for RG flow} cannot hold at arbitrarily long distances. For small interval sizes, and from the point of the UV theory, the deformation is relevant, produced by a current of weight $3/2$, and indeed expanding around the UV fixed point one finds power-law corrections starting with $\sim \lambda^4 (\Delta x)^2$, which become increasingly important for large $\Delta x\,$. On the other hand, for sufficiently large interval size (or large enough $\lambda$ for fixed interval size) one should instead find the IR theory result $S_{A} \to (c_{IR}/3)\log[\Delta x/a_{IR}]\,$, where  $c_{IR} = 4c_{UV}\,$. In the context of our prescription, the IR result is indeed recovered by taking the $\lambda \to \infty$ limit \textit{first}\footnote{We thank E. Perlmutter for pointing out that this order of limits yields the correct result in computations of scalar two-point functions on the RG flow background.}  in \eqref{RG flow loop 2}, and applying  \eqref{EE formula} with $\fudge=1$ as appropriate to the principal embedding theory. In particular, in this way one identifies
\begin{equation}
a_{IR} = \frac{e^{\rho_0/2}}{\lambda}a_{UV} = \frac{1}{e^{\rho_0 /2}\lambda}\,.
\end{equation}

The puzzling features of \eqref{EE for RG flow} may be due to the fact that we have not properly identified the cutoff. Since the UV and IR cutoffs are different, one expects the actual physical cutoff to interpolate between the two values and be a non-trivial function of $\lambda$. Moreover, since one also ends up in the IR regime for large $\Delta x$, the physical cutoff is presumably a non-trivial function of $\Delta x$ as well. In general, in AdS/CFT, we do not know how to relate bulk regularization (like choosing fixed $\rho_0$) to a particular regularization scheme in the boundary theory. Strong subadditivity is supposed to hold for a natural regularization in the boundary theory, but this may map to a complicated scheme from the bulk point of view. Perhaps the full result \eqref{RG flow loop 2} will give rise to a strongly subadditive entanglement entropy for a suitable choice of cutoff, but we leave a more detailed analysis of this interesting issue to future work.

%%%%%%%%%%%%%%%%%%%%%%%%%%%%%%%%%%%%%%%%%%%%%%%
\subsection{The charged black hole in the diagonal embedding}\label{sec:diag embedd bh}
%%%%%%%%%%%%%%%%%%%%%%%%%%%%%%%%%%%%%%%%%%%%%%%
We will now apply our formula to a finite temperature CFT state with a non-zero chemical potential for $U(1)$ charge turned on. The symmetry algebra in this case corresponds to two copies of a Virasoro algebra augmented by two copies of a $U(1)$ Kac-Moody algebra, and the $U(1)$ charge is furnished by the zero modes of the affine algebra. This can be realized from the bulk perspective by considering a black hole solution of the $N=3$ diagonal embedding theory in the truncation where the charged (spin-3/2) fields are turned off:
\begin{align}
a ={}& \left(\hat{\Lambda}^{+} - \mathcal{T}\hat{\Lambda}^{-} + jW_{0}\right)dx^{+} +\mu W_{0}\, dx^{-}
\\
\bar{a} ={}& -\left(\hat{\Lambda}^{-} - \overline{\mathcal{T}}\hat{\Lambda}^{+} + \overline{j}W_{0}\right)dx^{-} -\bar{\mu}W_{0}\, dx^{+}\,.
\end{align}

\noindent This solution corresponds to a BTZ black hole carrying $U(1)$ charge, and generalizes the non-rotating solution studied in \cite{Castro:2011fm}. The $sl(2,\mathds{R})$ generators in the diagonal embedding are given by \eqref{diag embed sl2 generators}, and $\mathcal{T}$ is the expectation value of the spectral flow-invariant combination of the stress tensor $T$ and the $U(1)$ current $U$ (we follow the conventions in \cite{Ammon:2011nk})
\begin{equation}\label{spectral flow invariant}
\mathcal{T} =\left\langle \frac{T}{k_{cs}} - \frac{3U^2}{4k_{cs}^2} \right\rangle = \left\langle \frac{6}{\hat{c}}\left(T -  \frac{27}{6\hat{c}}U^2\right)\right\rangle
\end{equation}

\noindent where $\hat{c} = 6k_{cs}$ is the central charge in the diagonal embedding (c.f. \eqref{central charge}), and the eigenvalue of $U$ is given by $(4k_{cs}/3)j$, with similar expressions in the barred sector.

Defining the matrices
\begin{equation}
h=2\pi \left(\tau a_{+} - \bar{\tau}a_{-}\right)\,,\qquad \bar{h} = 2\pi\left(\tau\bar{a}_{+}-\bar{\tau}\bar{a}_{-}\right),
\end{equation}

\noindent the smoothness conditions in the diagonal embedding, $\text{spec}\bigl(h\bigr) =\text{spec}\bigl(\bar{h}\bigr)= \text{spec}\left(2\pi i \hat{\Lambda}^{0}\right)$ (see section \ref{subsec:hsbh} for more details), can be recast as
\begin{equation}
\text{det}\left(h\right) = \text{det}\left(\bar{h}\right) =0\,,\qquad \text{Tr}\left[h^{2}\right] = \text{Tr}\left[\bar{h}^{2}\right]=-2\pi^{2}\,.
\end{equation}

\noindent The solution to these equations in the BTZ branch is
\begin{equation}
\tau = \frac{i}{2\sqrt{\mathcal{T}}}\,,\qquad \bar{\tau} = -\frac{i}{2\sqrt{\overline{\mathcal{T}}}}\,,\qquad \mu =\frac{\tau}{\bar{\tau}}\,j\,,\qquad \bar{\mu} =\frac{\bar{\tau}}{\tau}\,\overline{j} \,.
\end{equation}

\noindent Our general formula \eqref{our entropy formula} for the thermal entropy then yields the correct answer
\begin{equation}
S_{\textit{\scriptsize{thermal}}} =2\pi k_{cs} \left(\sqrt{\mathcal{T}}+ \sqrt{\overline{\mathcal{T}}}\right).
\end{equation}

Next, evaluating  \eqref{definition composite loop} and \eqref{EE formula} in the eight-dimensional adjoint representation (and with $\fudge = 2$), as appropriate to the diagonal embedding of the $N=3$ theory (c.f. table \ref{table reps}), we obtain
\begin{equation}
W_{Adj}(\Delta x) = e^{4\rho_0}\left(\frac{\sinh\left(\sqrt{\mathcal{T}}\Delta x\right)\sinh\left(\sqrt{\overline{\mathcal{T}}}\Delta x\right)}{\sqrt{\mathcal{T}\overline{\mathcal{T}}}}\right)^2 +\mathcal{O}\left(e^{2\rho_0}\right)
\end{equation}

\noindent and 
\begin{align}
S_{A} ={}&
k_{cs}\log\left[e^{2\rho_0}\frac{\sinh\left(\sqrt{\mathcal{T}}\Delta x\right)\sinh\left(\sqrt{\overline{\mathcal{T}}} \Delta x\right)}{\sqrt{\mathcal{T}\overline{\mathcal{T}}}}\right]
\\
={}&
\frac{\hat{c}}{6}\log\left[\frac{\beta_+\beta_{-}}{\pi^{2} a^{2}}\sinh\left(\frac{\pi \Delta x}{\beta_+}\right)\sinh\left(\frac{\pi \Delta x}{\beta_+}\right)\right]
\end{align}

\noindent where $a = e^{-\rho_0}$ is the cutoff and the inverse chiral temperatures $\beta_{\pm}=1/T_{\pm}$ are defined through $\mathcal{T} = \pi^2/\beta_{-}^{2}$ and $\overline{\mathcal{T}} = \pi^2/\beta_{+}^{2}\,$. 

Since the truncation of the diagonal embedding we are considering can be formulated as pure gravity coupled to Abelian gauge fields, we can in fact apply the R-T prescription to obtain the entanglement entropy for the dual of the charged BTZ solution. The corresponding calculation involves the length of geodesics on a standard BTZ black hole, with the only difference that the metric is written in terms of the expectation value of the spectral flow-invariant combination \eqref{spectral flow invariant} instead of that of the operators $L_0$, $\bar{L}_0\,$. According to \eqref{spectral flow invariant}, the result for the entanglement entropy should then agree with \eqref{BTZ ent ent 2d} upon replacing $T/k_{cs} \to \mathcal{T}$ and $\bar{T}/k_{cs} \to \overline{\mathcal{T}}\,$. It is reassuring to see that this is precisely the result we have obtained with our prescription.

%%%%%%%%%%%%%%%%%%%%%%%%%%%%%%%%%%%%%%%%%%%%%%%
\subsection{Higher spin black hole in the principal embedding}\label{subsec:hsbh}
%%%%%%%%%%%%%%%%%%%%%%%%%%%%%%%%%%%%%%%%%%%%%%%
We now discuss our main example, namely an application of our holographic entanglement entropy proposal to a CFT ensemble at finite temperature and finite higher spin charge. From the bulk perspective this is realized by considering the higher spin black hole solution constructed in \cite{Gutperle:2011kf,Ammon:2011nk}, which describes the CFT partition function at finite temperature and finite higher spin charge furnished by currents of weight $(3,0)$ (and $(0,3)$) \cite{Kraus:2011ds}.  We emphasize that the entanglement calculation on this background cannot be performed with any of the known holographic methods, so our result yields a non-trivial prediction. In the basis of generators $\{L_{i},W_{j}\}$ introduced in appendix \ref{sec:algebras}, the connections are of the form \eqref{general form connections} with $b = b(\rho) = e^{\rho \Lambda^{0}} = e^{\rho L_0}$ (principal embedding) and 
\begin{align}\label{highest weight gauge connection Gutperle Kraus}
a
={}&
\Bigl( L_{1}-\frac{2\pi \mathcal{L}}{k}\, L_{-1}-\frac{\pi \mathcal{W}}{2k}\,W_{-2}\Bigr) dx^{+}
\nonumber\\
&
+\mu \biggl(W_{2}+\frac{4\pi \mathcal{W}}{k}\,L_{-1}+\left( \frac{2\pi \mathcal{L}}{k}\right)^{2}\, W_{-2}-\frac{4\pi \mathcal{L}}{k}\,W_0 \biggr) dx^{-}\, ,
\\
\bar{a}
={}&
-\Bigl( L_{-1}-\frac{2\pi \bar{\mathcal{L}}}{k}\, L_{1}-\frac{\pi \bar{\mathcal{W}}}{2k}\,W_{2}\Bigr) dx^{-} 
\nonumber\\
&
-\bar{\mu} \biggl(W_{-2}+\frac{4\pi \bar{\mathcal{W}}}{k}\,L_{1}+\left( \frac{2\pi \bar{\mathcal{L}}}{k}\right)^{2}\, W_{2}-\frac{4\pi \bar{\mathcal{L}}}{k}\,W_0 \biggr) dx^{+}\,.
\label{highest weight gauge connection Gutperle Kraus 2}
\end{align}

\noindent Here, $\mathcal{W}$ and $\bar{\mathcal{W}}$ are the spin-3 charges, and $\mu$, $\bar{\mu}$ their conjugate chemical potentials. $\mathcal{L}$ and $\bar{\mathcal{L}}$ are related to the CFT stress tensor zero modes by $T = 2\pi \mathcal{L}\,$, $\bar{T} = 2\pi \bar{\mathcal{L}}\,$ (at least when the higher spin deformations are switched off; see \cite{Perez:2012cf,deBoer:2013gz,Compere:2013aa} for different definitions of the energy when $\mu$ and $\bar{\mu}$ are non-zero) . As before, $k$ is the level of the embedded $sl(2)$ theory, given by \eqref{sl2 k}, and related to the level $k_{cs}$ of the full theory via \eqref{kcs}: $k_{cs} = k/4\,$. Notice that the BTZ black hole connections \eqref{sl2r connections} are recovered by setting $\mathcal{W} = \bar{\mathcal{W}}=\mu=\bar{\mu}=0$. 

Let us say a few words about the smoothness properties of this solution. Analytically continuing to Euclidean time $t_{E}$ one can introduce complex coordinates $x^{+} = t +\varphi  \to z\,$, $x^{-}\to -\bar{z}\,$, and the topology of the solution is that of a solid torus. The boundary torus is defined by the identifications $z\simeq z + 2\pi \simeq z + 2\pi \tau$; for the BTZ solution in the $N=2$ theory $\tau_{BTZ} = i\beta\left(1+\Omega\right)/(2\pi )$, where $\beta$ and $\Omega$ are, respectively, the inverse temperature and angular velocity of the horizon ($\Omega$ is continued to purely imaginary values in order for the Euclidean section to be real).  The holonomies associated with the identification around the contractible cycle are
\begin{align}\label{thermal holonomies}
\mbox{Hol}_{\tau,\bar{\tau}}(A) 
=b^{-1}e^{h}\,b\,,\qquad
\mbox{Hol}_{\tau,\bar{\tau}}(\bar{A}) 
=
b\, e^{\bar{h}}b^{-1}\, ,
\end{align}
\noindent where the matrices $h$ and $\bar{h}$ are defined as
\begin{equation}
h = 2\pi\left(\tau a_{z} + \bar{\tau}a_{\bar{z}}\right)\,,\qquad \bar{h} = 2\pi\left(\tau \bar{a}_{z} + \bar{\tau}\bar{a}_{\bar{z}}\right).
\end{equation}
\noindent In \cite{Gutperle:2011kf,Ammon:2011nk} it was proposed that a gauge-invariant characterization of a regular black hole horizon in the higher spin theory is the requirement that the holonomies \eqref{thermal holonomies} are trivial, just as in the BTZ solution. In the principal embedding this condition can be rephrased as $\text{spec}(\tau a_{z} + \bar{\tau}a_{\bar{z}}) = \text{spec}(i\Lambda^0)\,$, and similarly for $\bar{h}\,$. For the $SL(3,\mathds{R})$ black hole solution at hand, this implies that the eigenvalues of $h$ and $\bar{h}$ in the fundamental representation are $(0,\pm2\pi i)$. Roughly speaking, the trivial holonomy requirement imposes relations between the charges and their conjugate potentials in a way consistent with thermodynamic equilibrium. 

In what follows we will focus on the non-rotating spin-3 black hole, obtained by setting 
\begin{equation}
\bar{\mathcal{L}} = \mathcal{L}\,,\qquad \bar{\mathcal{W}} = -\mathcal{W}\,,\qquad \bar{\mu} =-\mu\,.
\end{equation}
\noindent In the absence of rotation the modular parameter of the boundary torus is $\tau = -\bar{\tau} = i\beta/(2\pi)\,$, where $\beta$ is the inverse temperature, and the holonomy matrices become simply $h = 2\pi \tau \, a_{t} \,$, $\bar{h} = 2\pi \tau\, \bar{a}_{t}\,$. The smoothness conditions then boil down to the requirement that the holonomy around the Euclidean time circle is trivial. In principle there exist multiple solutions to the holonomy equations, corresponding to different thermodynamic phases (in the $N=3$ case, these phases were explored in \cite{David:2012iu}). Here we will concentrate in the BTZ branch, defined by the requirement that we recover the BTZ results when all the higher spin charges and chemical potentials are switched off. The holonomy conditions for the spin-3 black hole  \cite{Gutperle:2011kf,Ammon:2011nk} can be solved explicitly in the non-rotating limit: in the BTZ branch one finds
\begin{equation}\label{solution smoothness GK}
\mathcal{W} = \frac{4(C-1)}{C^{3/2}}\mathcal{L}\sqrt{\frac{2\pi \mathcal{L}}{k}}\,,\qquad \mu = \frac{3\sqrt{C}}{4(2C-3)}\sqrt{\frac{k}{2\pi \mathcal{L}}}\,,\qquad \tau = \frac{i\left(2C-3\right)}{4\left(C-3\right)\sqrt{1-\frac{3}{4C}}}\sqrt{\frac{k}{2\pi \mathcal{L}}}\,,
\end{equation}

\noindent where $C > 3$ and $C =\infty$ at the BTZ point. Since $\tau= i\beta/(2\pi)$ in the non-rotating case, we notice that $C$ can be thought of as parameterizing the dimensionless ratio $\frac{\mu}{\beta}$:
\begin{equation}\label{mu over beta}
\frac{\mu}{\beta} = \frac{3}{4\pi}\frac{(C-3)\sqrt{4C-3}}{(3-2C)^2}\,.
\end{equation}

Having solved the smoothness conditions, we can now go back to Lorentzian signature and consider the solution with a non-compact boundary spatial coordinate. This is, from the dual CFT perspective we consider a finite temperature system on the infinite line, with a non-zero chemical potential for spin-3 charge. From the general discussion in section \ref{subsec:choice of rep}, we know that the appropriate representation $\mathcal{R}$ for the $N=3$ theory in the principal embedding is the 8-dimensional adjoint representation (and with $\fudge =1$). First, we notice that the eigenvalues of $a_{x}$ in the adjoint representation are $(\pm 0, \pm \lambda^{(1)}_{Adj},\pm\lambda^{(2)}_{Adj}, \pm \lambda^{(3)}_{Adj})$ with
\begin{align}\label{aphi eigen}
\lambda^{(1)}_{Adj} ={}& 4\sqrt{\frac{2\pi \mathcal{L}}{k}}\,\frac{\sqrt{1-\frac{3}{4C}}}{1-\frac{3}{2C}}=\frac{4\pi}{\beta}\frac{1}{\left(1-\frac{3}{C}\right)}
\\
\lambda^{(2)}_{Adj} ={}& 2\sqrt{\frac{2\pi \mathcal{L}}{k}}\left(\frac{3}{\sqrt{C}}+ \frac{\sqrt{1-\frac{3}{4C}}}{1-\frac{3}{2C}}\right)=  \frac{2\pi}{\beta}\frac{1}{\left(1-\frac{3}{C}\right)}\left(\frac{3}{\sqrt{C}}\frac{\left(1-\frac{3}{2C}\right)}{\sqrt{1-\frac{3}{4C}}} +1\right) 
\\
\lambda^{(3)}_{Adj} ={}& 2\sqrt{\frac{2\pi \mathcal{L}}{k}}\left(\frac{3}{\sqrt{C}} - \frac{\sqrt{1-\frac{3}{4C}}}{1-\frac{3}{2C}}\right) = \frac{2\pi}{\beta}\frac{1}{\left(1-\frac{3}{C}\right)}\left(\frac{3}{\sqrt{C}}\frac{\left(1-\frac{3}{2C}\right)}{\sqrt{1-\frac{3}{4C}}} -1\right).
\label{aphi eigen 2}
\end{align}

\noindent As usual, these eigenvalues correspond to the pairwise difference of the eigenvalues in the fundamental representation. Evaluating the leading term in \eqref{definition composite loop} as $\rho_0 \to \infty$ we obtain
\begin{align}\label{adjoint loop}
W_{Adj}(P,Q)
\simeq{}&
 \left(\frac{\beta}{\pi a}F\left(\frac{\Delta x}{\beta},C\right) \right)^8
\end{align}

\noindent where $a = e^{-\rho_0}$ as before and the auxiliary function $F\left(\frac{\Delta x}{\beta},C\right)$ is defined through\footnote{Note that the combinations $\lambda^{(i)}_{Adj}\Delta x$ ($i=1,2,3$) depend on the dimensionless quantities $C$ and $ \frac{\Delta x}{\beta }$ only.}
\begin{align}\label{definition F}
F^{8}\left(\frac{\Delta x}{\beta},C\right) ={}&
\frac{\left(1-\frac{3}{C}\right)^4\left(1-\frac{3}{4C}\right)^{2}}{\left(1-\frac{3}{2C}\right)^8}\Biggl[\frac{3}{8}  +\frac{1}{8} \sqrt{1-\frac{3}{4C}}\left(\frac{3}{\sqrt{C}} - 2\sqrt{1-\frac{3}{4C}}\right)\cosh\left(\lambda_{Adj}^{(2)}\Delta x\right)
 \nonumber\\
 &
+ \frac{1}{8} \left(1-\frac{3}{C}\right)\cosh\left(\lambda^{(1)}_{Adj}\Delta x\right)-\frac{1}{8} \sqrt{1-\frac{3}{4C}}\left(\frac{3}{\sqrt{C}} + 2\sqrt{1-\frac{3}{4C}}\right)\cosh\left(\lambda_{Adj}^{(3)}\Delta x\right) \Biggr]^2.
\end{align}

\noindent Using \eqref{central charge} in the principal embedding we get $c = 24k_{cs}\,$, and \eqref{EE formula} then yields
\begin{equation}\label{higher spin bh EE}
S_{A} = \frac{c}{3}\log\left[\frac{\beta}{\pi\,a}\left| F\left(\frac{\Delta x}{\beta},C\right) \right| \right].
\end{equation}

\noindent In what follows we will study several limits of our general result \eqref{definition F}-\eqref{higher spin bh EE}.

\subsubsection{BTZ limit and perturbative corrections}
As a first check of our result, we can easily see that it reduces to the universal finite temperature entanglement when the spin-3 charge and chemical potential are switched off. To this end it suffices to notice that in the BTZ limit given by $C \to \infty$ with $\mathcal{L}$ finite (so that $\beta$ is fixed and $\mu \to 0$, $\mathcal{W}\to 0$) we have
\begin{equation}
\mbox{BTZ limit}:\quad \lambda^{(1)}_{Adj} = 2\lambda^{(2)}_{Adj} = -2\lambda^{(3)}_{Adj}  = \frac{4\pi}{\beta} \quad \Rightarrow\quad  F\left(\frac{\Delta x}{\beta},C \to \infty\right) = \sinh\left( \frac{\pi \Delta x}{\beta}\right),
\end{equation}

\noindent and \eqref{higher spin bh EE} immediately reduces to the non-rotating limit of \eqref{BTZ ent ent 2d} (namely \eqref{finite temp ent ent}). More generally, expanding the result perturbatively in $\mu \to 0$  with the inverse temperature $\beta$ held fixed our general expression \eqref{higher spin bh EE} yields
\begin{align}\label{pert expansion}
S_{A} &\xrightarrow[\mu \to 0]{} 
 \frac{c}{3}\log\left[\frac{\beta}{a\pi}\sinh\left(\frac{\pi\Delta x}{\beta}\right)\right] + \frac{c}{18}\left(\frac{\pi \mu}{\beta}\right)^2\text{csch}^4\left(\frac{\pi\Delta x}{\beta}\right)\biggl[-3 -5\cosh\left(4\frac{\pi\Delta x}{\beta}\right)
\nonumber\\
&
+ 8\left(1-12\left(\frac{\pi\Delta x}{\beta}\right)^2\right)\cosh\left(2\frac{\pi\Delta x}{\beta}\right) + 16\frac{\pi\Delta x}{\beta}\left(\sinh\left( 2\frac{\pi\Delta x}{\beta}\right)+\sinh \left(4\frac{\pi\Delta x}{\beta}\right)\right)\biggr]
\nonumber\\
& + \mathcal{O}(\mu^4)
\end{align}

\noindent  While reproducing our full non-perturbative result \eqref{definition F}-\eqref{higher spin bh EE} with an independent CFT calculation is presumably very hard, it is plausible that an expansion such as \eqref{pert expansion} could be checked on a term-by-term basis by perturbatively evaluating the two-point function of twist fields in the presence of a deformation by the higher spin current.   

\subsubsection{Extensive (high temperature) limit}
By construction, we know that in the high temperature limit the result \eqref{higher spin bh EE} will reproduce the appropriate thermal entropy in an ensemble with higher spin charge. It is however instructive to see explicitly how this result is recovered. We first notice that
\begin{equation}
\lambda^{(1)}_{Adj} > \lambda^{(2)}_{Adj} >\lambda^{(3)}_{Adj} \qquad \forall \,\,\,C > C_{0} \equiv \frac{3}{8}\left(9 +\sqrt{33}\right) \simeq 5.53
\end{equation}

\noindent so, starting in a neighborhood of the BTZ point, it is always possible to order the eigenvalues. There is eigenvalue crossing at the value $C_0$ introduced above, and it is conceivable that this indicates a phase transition along the lines studied in \cite{David:2012iu}, but here we focus on the $C > C_0$ portion of the BTZ branch and neglect this possibility. Then, taking $\Delta x$ very large and looking at the extensive contribution to the entanglement entropy, we find that $\lambda^{(1)}_{Adj}$ is the dominant eigenvalue, and \eqref{adjoint loop} reduces to
\begin{equation}\label{adjoint thermal loop}
W_{Adj}(P,Q)
\xrightarrow[\Delta x\, \gg \,\beta]{}
\left(\frac{\beta}{\pi a}\frac{\left(1-\frac{3}{4C}\right)^{1/4}\left(1-\frac{3}{C}\right)^{3/4}}{2\left(1-\frac{3}{2C}\right)}\right)^8 \exp\left(2\lambda^{(1)}_{Adj}\,\Delta x\right),
\end{equation}

\noindent and subtracting the UV divergence we find
\begin{equation}
S_{A} \xrightarrow[\Delta x \,\gg\, \beta]{} k_{cs}\log \exp\left(2\lambda^{(1)}_{Adj}\,\Delta x\right) = 2\sqrt{2\pi k \mathcal{L}}\,\frac{\sqrt{1-\frac{3}{4C}}}{1-\frac{3}{2C}}\,     \Delta x = s_{\textit{\scriptsize{thermal}}}\, \Delta x\,,
\end{equation}

\noindent where the thermal entropy density $s_{\textit{\scriptsize{thermal}}}$ is defined as before:
\begin{equation}
s_{\textit{\scriptsize{thermal}}} = \frac{S_{\textit{\scriptsize{thermal}}}}{2\pi} =2\sqrt{2\pi k \mathcal{L}}\,\frac{\sqrt{1-\frac{3}{4C}}}{1-\frac{3}{2C}}\,.
\end{equation}

\noindent This is consistent with the thermal entropy $S_{\textit{\scriptsize{thermal}}}$ of the spin-3 black hole as computed in \cite{Perez:2013xi,Campoleoni:2012hp}. We point out that a different result for the thermal entropy was given in \cite{Gutperle:2011kf,Ammon:2011nk}. In \cite{deBoer:2013gz} we explained how these different results correspond to different choices of boundary conditions, and we will further elaborate on this delicate point in the discussion section.

\subsubsection{Zero-temperature limit}
Let us now focus on the zero-temperature of our result. From \eqref{mu over beta} we see that, for fixed $\mu$, in the allowed range of $C$ the low temperature limit $\beta \to \infty$ can be achieved with $C\to \infty$ or $C \to 3$. Let us first focus in the $C \to \infty$ regime. Unlike the BTZ limit, in this case we scale $\mathcal{L} \sim 1/C \to 0$, so that $\mu$ is finite and $\mathcal{W} \to 0$, obtaining
\begin{equation}\label{zero temp hs bh result}
S_{A} \simeq \frac{c}{3}\log\left[\frac{\Delta x}{a}\left(1-16\frac{\mu^{2}}{\Delta x^{2}}\right)^{1/4}\right].
\end{equation}

\noindent If we think of the black hole solution as a finite temperature generalization of the RG flow studied in section \ref{RG flow}, one could have anticipated that the result cannot hold at arbitrary short distances. Indeed, the above result possesses features reminiscent of those of \eqref{EE for RG flow} (although the latter must be interpreted from the perspective of the UV theory, while \eqref{zero temp hs bh result} is a deformation of the IR theory). As a further check we expand
the above result in the deformation parameter $\mu$ to obtain
\begin{equation}
S_{A} \simeq
 \frac{c}{3}\log\left[\frac{\Delta x}{a}\right] - \frac{4c}{3}\biggl(\frac{\mu}{\Delta x}\biggr)^2 +\mathcal{O}\left(\frac{\mu^{4}}{\Delta x^4}\right)
\end{equation}

\noindent The first term is just the familiar zero-temperature single-interval entanglement entropy in the IR theory, while the first correction scales as $\left(\Delta x\right)^{-2}$ with the interval size, consistent with the fact that the operator responsible for the perturbation has conformal dimension $\Delta_{\hat{W}} = 3\,$ at the IR fixed point \cite{Cardy:2010zs}.

If we instead take the zero temperature limit by letting $C\to 3$, so that $\beta \to \infty$ with $\mu$ and $\mathcal{W}$ finite (i.e. the extremal black hole limit), we obtain
\begin{equation}
S_{A} \xrightarrow[\beta \to \infty]{\mu, \mathcal{W} \text{ finite}} \frac{c}{3}\log\left[\frac{2}{3}\frac{\mu}{a}\left(\left(\frac{3\Delta x}{\mu}\right)^2 +8\cosh\left(\frac{3\Delta x}{\mu}\right)-8 + 4\left(\frac{3\Delta x}{\mu}\right)\sinh\left(\frac{3\Delta x}{\mu}\right)\right)^{1/4}\right].
\end{equation}

\noindent The fact that this expression does not have a smooth $\mu \to 0$ limit can be understood as follows: at zero temperature, the connection corresponding to the (extremal) BTZ black hole cannot be diagonalized. Therefore, while the finite-temperature higher spin black hole connects smoothly to a finite-temperature BTZ black hole as we turn off the higher spin charge, it is not clear what the appropriate notion of smoothness is for the extremal black hole. It would be of interest to discuss extremal higher spin black holes in general and to study their properties. 

\subsubsection{Short distance behavior and the UV cutoff}
\begin{figure}[h!]
\begin{center}
\includegraphics[width=3.1in]{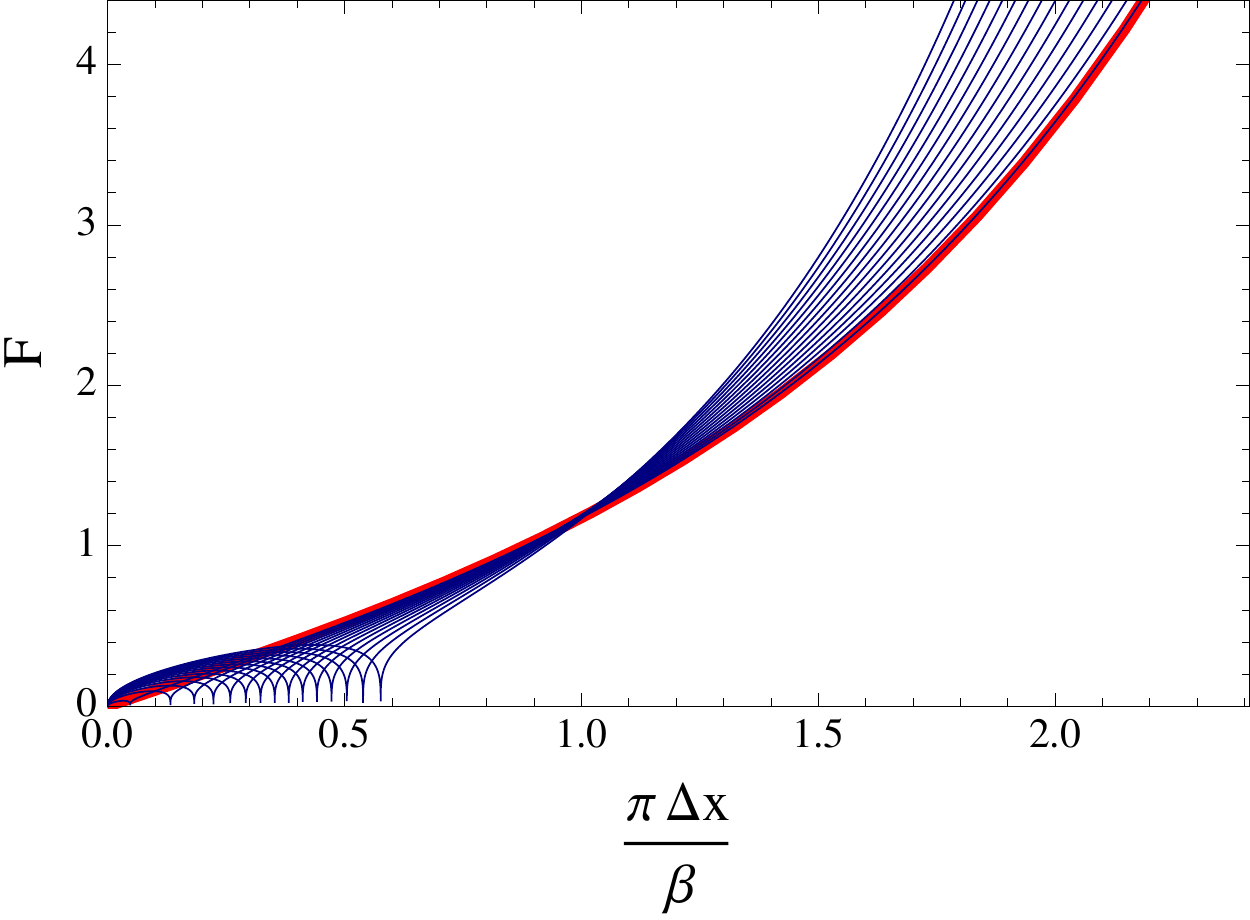}\quad
\includegraphics[width=3.1in]{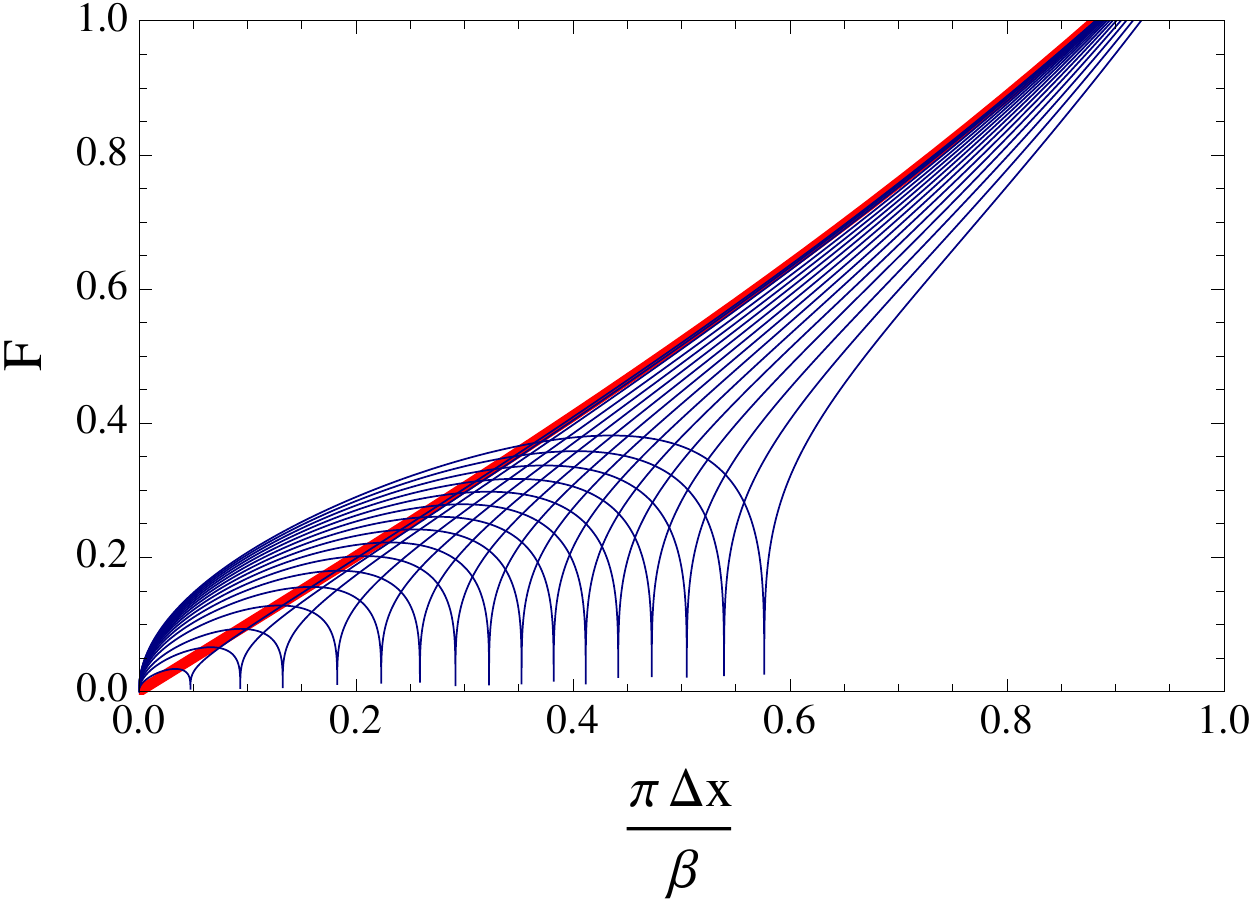}
\end{center}
\caption{Left: $F\left(\frac{\Delta x}{\beta},C\right)$ as a function of $\pi\frac{\Delta x}{\beta}$ for fixed $\mu/\beta$ (fixed $C$). The red curve corresponds to the result in the absence of higher spin charges, $F\left(\frac{\Delta x}{\beta},\infty\right)=\sinh\left(\pi\frac{\Delta x}{\beta}\right)$. The blue curves correspond to the higher spin result for different values of $C \in \left[10,1000\right]$ ($\mu/\beta \in \left[0.0038,0.035\right]$). Right: Zoom into the short-distance regime suggesting a redefinition of the cutoff.}
   \label{fig:sing}
\end{figure}

\begin{figure}[h!]
\vspace{1.3cm}
\centering
\includegraphics[width=10cm]{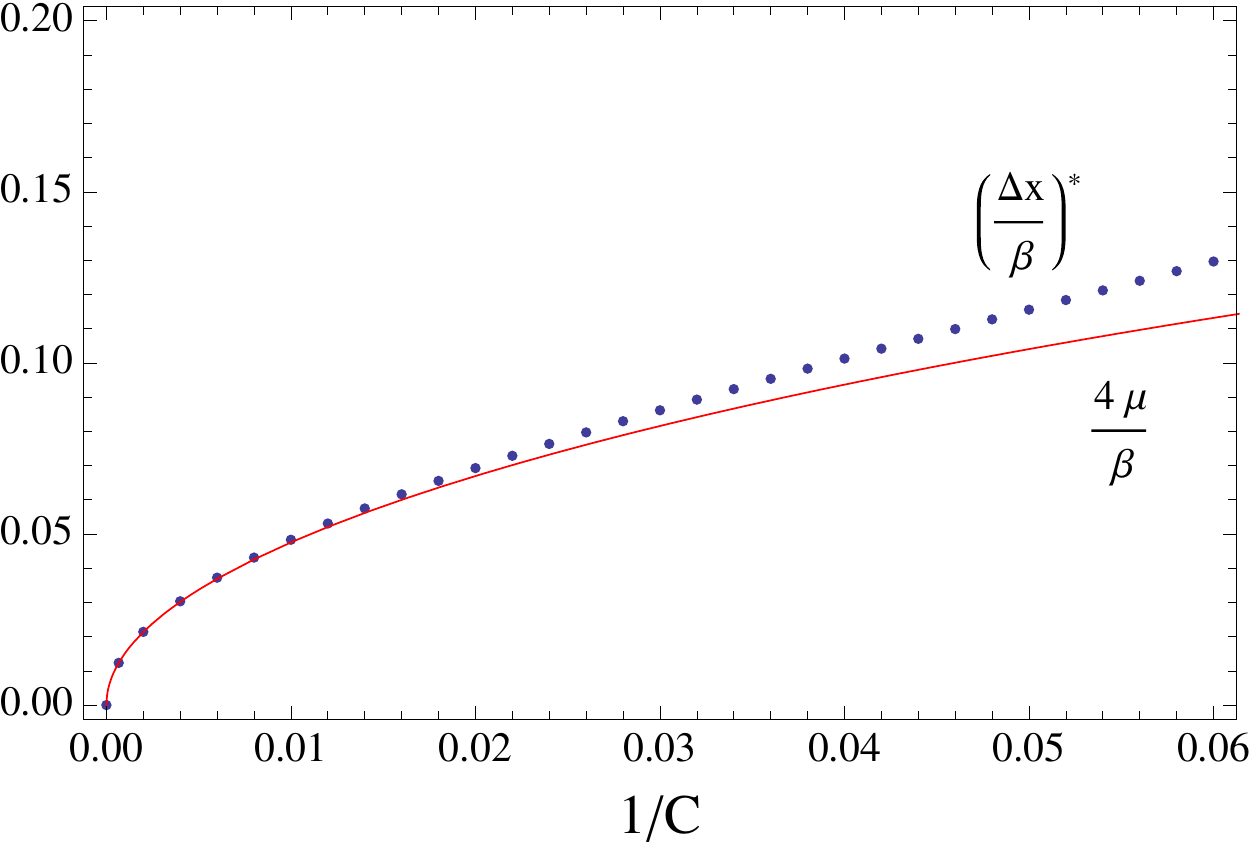}
\caption{Dots: Numerically-determined values of $\Delta x/\beta$ at which the entanglement expression breaks down, as a function of $1/C$. Solid line: $\mu /\beta$  as a function of $1/C$.}
\label{fig:breaking}
\end{figure}

A numerical analysis of our result \eqref{higher spin bh EE} reveals that the function $F\left(\frac{\Delta x}{\beta},C\right)$ (and hence $W_{Adj}$) approaches zero for a non-zero value of $\Delta x/\beta$ which depends on $C$ (equivalently, on $\mu/\beta$). This behavior is illustrated in figure \ref{fig:sing}. We will denote the critical value of $\Delta x/\beta$ by $(\Delta x/\beta)^*$, i.e. $F\left((\Delta x/\beta)^*,C\right)=0$. In figure \ref{fig:breaking} we have plotted the numerically-determined value $\left(\Delta x/\beta\right)^*$ versus the corresponding value of $\mu /\beta$, both as a function of the dimensionless parameter $1/C$.

\begin{figure}[h!]
\vspace{1cm}
\centering
\includegraphics[width=10cm]{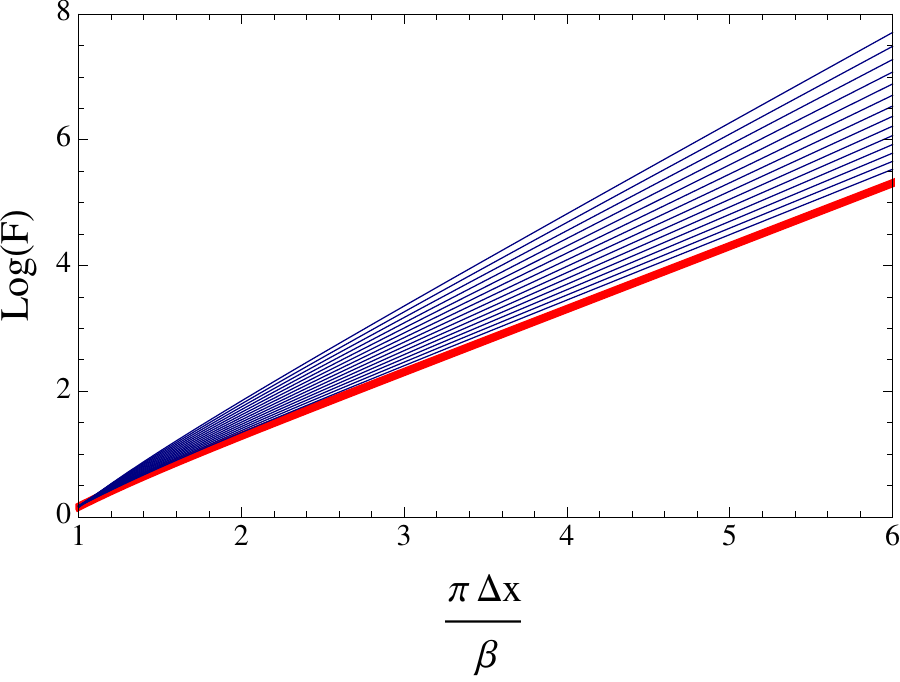}
\caption{$\log\left[F\left(\frac{\Delta x}{\beta},C\right)\right]$ as a function of $\pi\frac{\Delta x}{\beta}$ for different fixed values of $C$ (equivalently $\mu/\beta$) and $(\Delta x/\beta) > (\Delta x/\beta)^*$. The red curve shows the result in the absence of higher spin charge. The blue curves show the higher spin result for different values of $\mu/\beta$.}
\label{fig:Reg}
\end{figure}

 For small $\mu$ (i.e. close to the BTZ point $1/C\to 0$) we observe that the critical value $\Delta x^*$ is very well approximated by $\Delta x^* \simeq 4\mu$. It is then plausible that the breakdown of the result for small values of  $\Delta x \simeq \mu$ is indicating the necessity to redefine the UV cutoff due to the effect of the irrelevant perturbation. This interpretation would be consistent with the fact that no such singularities were observed in the calculation involving the diagonal embedding black hole (c.f. section \ref{sec:diag embedd bh}), where the current sourced by the $U(1)$ chemical potential is relevant. Moreover, as shown in figure \ref{fig:Reg}, for $(\Delta x/\beta) > (\Delta x/\beta)^*$ our result for the entanglement behaves in a way consistent with regularity and strong subadditivity.

Besides the possibility that we may need to redefine the cutoff as in the RG flow example in section \ref{RG flow}, it is also feasible that the theory is ill-defined at short distances, which is the naive expectation in the presence of irrelevant interactions. In such case one should first find a suitable UV completion in order to define entanglement entropy at short distances.

%%%%%%%%%%%%%%%%%%%%%%%%%%%%%%%%%%%%%%%%%%%%%%%%%%%%%%%%%%%%%%%%%%%%%%%%%%%%%%%%
\section{Discussion and outlook} \label{sec: conclusions}
%%%%%%%%%%%%%%%%%%%%%%%%%%%%%%%%%%%%%%%%%%%%%%%%%%%%%%%%%%%%%%%%%%%%%%%%%%%%%%%%
Inspired by the recent developments in three-dimensional higher spin holography, we have studied the problem of computing entanglement entropy in $2d$ CFTs with $\mathcal{W}_{N}$ symmetries using holographic techniques. In particular, we introduced a bulk functional, c.f. \eqref{EE formula}, that captures the entanglement entropy in  in the absence of higher spin charges, where universal field-theoretical results are available, and admits an immediate generalization to the higher spin setup, even in the presence of non-trivial higher spin chemical potentials corresponding to CFTs perturbed by higher spin currents. Let us summarize some of the features of this functional:
\begin{itemize}
\item It is written solely in terms of Wilson lines, as appropriate to the topological character of the bulk Chern-Simons theory. Moreover, it is manifestly path-independent when the connections satisfy the equations of motion, i.e. it depends only on the positions of the boundary points and the homotopy class of the path. 

\item For a single interval in one spatial dimension, it correctly reproduces the CFT entanglement entropy in all the cases where there is an independent field-theoretical understanding of the result, including situations with finite temperature and finite angular momentum.

\item By construction, the correct thermal entropy is recovered in the limit in which the von Neumann entropy becomes extensive, even in the presence of higher spin charges and chemical potentials.
\end{itemize}

In the absence of explicit field-theoretical calculations of entanglement entropy in the presence of non-trivial higher spin charges, the above list provides evidence in favor of the holographic entanglement entropy interpretation. Let us however point out that other definitions are possible; to illustrate this point, consider a ``holomorphically-factorized" version of \eqref{definition composite loop}:
\begin{align}\label{holo definition composite loop}
 W^{holo}_{\mathcal{R}}(P,Q)
\equiv{}&
\mbox{Tr}_{\mathcal{R}}\Biggl[\mathcal{P}\exp\left(\int_{Q}^{P}\bar{A}_{-}\,dx^-\right)\,\mathcal{P}\exp\left(\int_{P}^{Q}A_+\,dx^+\right)\Biggr].
\end{align}

\noindent Since the truncated connections $\bar{A}_{-}\,dx^{-}$ and $A_{+}\,dx^{+}$ are in general not flat by themselves,  an immediate shortcoming of this expression is that it is not in general path-independent, as opposed to \eqref{definition composite loop}. Despite this fact, let us momentarily focus on connections whose components are independent of the boundary coordinates $x^\pm\,$, so that \eqref{holo definition composite loop} is well-defined. Computing \eqref{holo definition composite loop} for the charged, non-rotating, spin-3 black hole studied in section \ref{subsec:hsbh} we obtain (using the adjoint representation as appropriate to the principal embedding in the $N=3$ theory)
\begin{align}
W^{holo}_{Adj}(P,Q) 
={}&
 \left(\frac{8e^{4\rho_{0}}}{\tilde{\lambda}_{1}\tilde{\lambda}_{2}\tilde{\lambda}_{3}}\right)^{2}\left(\frac{\tilde{\lambda}_{1}^{2}-\tilde{\lambda}_{2}\tilde{\lambda}_{3}}{\tilde{\lambda}_{1}\tilde{\lambda}_{2}\tilde{\lambda}_{3}} + \frac{\cosh(\tilde{\lambda}_{1}\Delta x)}{\tilde{\lambda}_{1}}-\frac{\cosh(\tilde{\lambda}_{2}\Delta x)}{\tilde{\lambda}_{2}}-\frac{\cosh(\tilde{\lambda}_{3}\Delta x)}{\tilde{\lambda}_{3}}\right)^{2}\,,
\end{align}

\noindent where  $(\pm 0, \pm \tilde{\lambda}_{1},\pm\tilde{\lambda}_{2}, \pm \tilde{\lambda}_{3})$ are the eigenvalues of $a_{+}$ (as opposed to those of $a_{x}$) in the adjoint representation, with
\begin{align}\label{eigenvalues ap}
 \tilde{\lambda}_{1} ={}&
4\sqrt{\frac{2\pi \mathcal{L}}{k}}\sqrt{1 - \frac{3}{4C}}\,,
\\
 \tilde{\lambda}_{2} ={}&
2\sqrt{\frac{2\pi \mathcal{L}}{k}}\left(\sqrt{1 - \frac{3}{4C}}-\frac{3}{2\sqrt{C}}\right),\\
   \tilde{\lambda}_{3} ={}&
2\sqrt{\frac{2\pi \mathcal{L}}{k}}\left(\sqrt{1 - \frac{3}{4C}}+\frac{3}{2\sqrt{C}}\right).
\end{align}

\noindent If one now defines $S_{A}^{holo} = k_{cs}\log W^{holo}(P,Q)\,$, the corresponding $S_{A}^{holo}$ is strongly subadditive as well, and reduces to the universal result \eqref{finite temp ent ent} when the higher spin charge is switched off (namely at the BTZ point $C \to \infty$). 
In the large $\Delta x$ limit, $\tilde{\lambda}_{1}$ is the dominant eigenvalue for all allowed values of $C$, and upon subtraction of the usual UV divergence one finds
\begin{equation}\label{spin 3 bh GK entropy}
S^{holo}_{A} \xrightarrow[\Delta x \,\gg\, \beta]{} k_{cs}\log  \exp\left(2\tilde{\lambda}_{1}\Delta x\right) =2\sqrt{2\pi k \mathcal{L}}\,\sqrt{1-\frac{3}{4C}}\,\Delta x = s^{holo}_{\textit{\scriptsize{thermal}}}\, \Delta x \,.
\end{equation}

\noindent where the holomorphic thermal entropy density $s^{holo}_{\textit{\scriptsize{thermal}}}$ is defined as 
\begin{equation}\label{spin 3 bh GK entropy 2}
s^{holo}_{\textit{\scriptsize{thermal}}} = \frac{S^{holo}_{\textit{\scriptsize{thermal}}}}{2\pi } =2\sqrt{2\pi k \mathcal{L}}\sqrt{1-\frac{3}{4C}}\,.
\end{equation}

\noindent This result is consistent with the thermal entropy of the spin-3 black hole as computed in \cite{Gutperle:2011kf,Ammon:2011nk} (see \cite{Kraus:2013fk} also). As we showed in \cite{deBoer:2013gz}, this result is obtained from the bulk on-shell action with ``holomorphic" boundary conditions, and it is consistent with a thermal entropy given by (in the principal embedding)
\begin{equation}\label{our holomorphic Cardy formula v2}
S^{holo}_{\textit{\scriptsize{thermal}}}  = 2\pi k_{cs}\langle \vec{\lambda}_{+}-\vec{\overline{\lambda}}_{-},\rho \rangle\,,
\end{equation}

\noindent where $\vec{\lambda}_+$ is the weight vector dual to the Cartan algebra element that is conjugate to $a_{+}$ (and similarly for $\vec{\overline{\lambda}}_-$). Equation \eqref{our holomorphic Cardy formula v2} should be contrasted with the ``canonical" entropy \eqref{our Cardy formula v2}, that involves the eigenvalues of $a_{x}$ rather than those of $a_{+}\,$. It is worth pointing out that the result \eqref{spin 3 bh GK entropy 2} for the thermal entropy of the spin-3 black hole has been found to be in agreement with independent CFT calculations of the partition function \cite{Kraus:2011ds,Gaberdiel:2012yb}. It is interesting that the canonical result \eqref{our Cardy formula v2} is recovered form the somewhat more natural form \eqref{definition composite loop} rather than the holomorphically factorized version \eqref{holo definition composite loop}, which is harder to justify a priori. Ultimately, we expect the entanglement calculations presented here to shed some light on the nature of the canonical result for the thermal entropy, whose detailed CFT interpretation is still lacking (see \cite{Compere:2013aa} for a recent discussion of charges and asymptotic symmetries in the presence of higher spin chemical potentials).

A comparison against independent field-theoretical calculations of entanglement entropy in the presence of higher spin sources would of course be the litmus test for our prescription. Such calculations are however difficult, inasmuch as they presume detailed knowledge of correlation functions of twist operators in the presence of deformations by (irrelevant) higher spin operators. To our knowledge, no explicit results are known presently. It is however conceivable that a matching could be achieved for specific instances of the bulk theory based on the so-called $hs[\lambda]$ algebra, which would require adapting our prescription to an infinite-dimensional gauge algebra. We expect the latter obstacle to be of a purely technical nature. 

There are several other interesting directions to explore. We would obviously like to have a better understanding of the possible breakdown of strong subadditivity for short distances in these higher spin theories, and to come up with a general analytic proof of strong subadditivity at long distances. One could also try to construct extensions of our proposal to include the so-called R\'enyi entropies that feature prominently in the CFT calculations of entanglement via the replica trick. Indeed, one can in principle obtain $\text{Tr}[\rho_{A}^{n}]$ holographically by computing the Chern-Simons partition function for bulk solutions that asymptote to boundary geometries that are branched covers of the original solution with branch points at the endpoints of the interval, as it was recently done in \cite{Faulkner:2013yia} for the standard gravity case. Alternatively, one could try to directly compute the correlation functions of twist fields via holography. However, since twist fields are not included in the Chern-Simons theory, one would probably have to couple matter fields to it, which would require us to use the full $3d$ Vasiliev theory. In \cite{Kraus:2012uf} such two-point functions were computed and the results have a striking similarity to our expression for the entanglement entropy. Furthermore, the peculiar difference between the canonical and holomorphic formulations of both the ordinary as well as the entanglement entropy is clearly crying out for a better understanding, as does the question of whether theories with sources for the higher spin currents are non-perturbatively well defined. We leave these very interesting problems for future work.

%%%%%%%%%%%%%%%%%%%%%%%%%%%%%%%%%%%%%%%%%%%%%%%%%%%%%%%%%%%%%%%%%%%%%%%%%%%%%%%%
%%%%%%%%%%%%%%%%%%%%%%%%%%%%%%%%%%%%%%%%%%%%%%%%%%%%%%%%%%%%%%%%%%%%%%%%%%%%%%%%
\vskip 1cm
\centerline{\bf Acknowledgments}
We are specially grateful to Matt Headrick, Eric Perlmutter and Mukund Rangamani for valuable discussions and detailed comments on an earlier version of the manuscript. We also thank Marco Baggio, Max Ba\~nados, Alejandra Castro, Jean-S\'ebastien Caux, Geoffrey Comp\`ere, Michal Heller, Veronika Hubeny, Robert Konik, Rob Leigh, Tatsuma Nishioka, Andrei Parnachev, Wei Song and Ari Turner for helpful conversations. This work is part of the research programme of the Foundation for Fundamental Research on Matter (FOM), which is part of the Netherlands Organization for Scientific Research (NWO).

%%%%%%%%%%%%%%%%%%%%%%%%%%%%%%%%%%%%%%%%%%%%%%%%%%%%%%%%%%%%%%%%%%%%%%%%%%%%%%%%
%%%%%%%%%%%%%%%%%%%%%%%%%%%%%%%%%%%%%%%%%%%%%%%%%%%%%%%%%%%%%%%%%%%%%%%%%%%%%%%%
\appendix
%%%%%%%%%%%%%%%%%%%%%%%%%%%%%%%%%%%%%%%%%%%%%%%%%%%%%%%%%%%%%%%%%%%%%%%%%%%%%%%%
\section{First order formalism and the Chern-Simons formulation of $3d$ Einstein gravity\label{sec: vielbein formalism}}
%%%%%%%%%%%%%%%%%%%%%%%%%%%%%%%%%%%%%%%%%%%%%%%%%%%%%%%%%%%%%%%%%%%%%%%%%%%%%%%%
We denote local Lorentz indices by latin characters $a,b,\ldots$ and spacetime indices by Greek letters $\mu,\nu,\ldots\,$. The basic variables in the first order formalism are the vielbein\footnote{More properly, in the three-dimensional case we should refer to it as \textit{dreibein} or \textit{triad}.} $e^{a} = e^{a}_{\hphantom{a}\mu}dx^{\mu}$, such that $ds^{2}=g_{\mu\nu}\,dx^{\mu}\otimes dx^{\nu} = \eta_{ab}\,e^{a}\otimes e^{b}$ with $\eta = \mbox{diag}(-1,1,1)$, and the spin connection $\omega^{a}_{\hphantom{a}b} = \omega^{a}_{\hphantom{a}cb}e^{c}\,$. In this language, the condition of metric-compatibility on the connection reads $\omega_{ab} = -\omega_{ba}\,$; hence, $\omega_{ab}$ can be thought of as a one-form valued in antisymmetric 3$\times$3 matrices.

Int 3$d$ it is convenient to dualize the spin connection and define
\begin{equation}\label{definition dual spin connection}
\omega^{a} \equiv \frac{1}{2}\epsilon^{abc}\omega_{bc}\qquad  \Leftrightarrow\qquad \omega_{ab} = -\epsilon_{abc}\,\omega^{c}\, ,
\end{equation}
\noindent where $\epsilon_{abc}$ are the components of the Levi-Civita tensor in the local Lorentz frame.  We adopt the convention $\epsilon_{012}=-1 \Rightarrow \epsilon^{012} = +1$. In terms of $\omega^{a}$, Cartan's structure equations read
\begin{align}
R^{a} 
={}& d\omega^{a} + \frac{1}{2}\epsilon^{a}_{\hphantom{a}bc}\,\omega^{b}\wedge \omega^{c}
\\
T^{a} 
={}&
 de^{a} -\epsilon^{a}_{\hphantom{a}bc}\,\omega^{c}\wedge e^{b}\,,
\end{align}
\noindent where $R^{a}\equiv \frac{1}{2}\epsilon^{abc}R_{bc}$ is the dual of the standard curvature two-form $R_{ab}\,$, and $T^{a}$ is the torsion. Next, we introduce $A =  \omega +\frac{e}{\ell}$ and $ \bar{A} =  \omega - \frac{e}{\ell}\,$, where $\omega = \omega^{a}J_{a}\,$, $e = e^{a}J_{a}$. Defining the Chern-Simons form
\begin{equation}
CS(A) = A\wedge dA + \frac{2}{3}A\wedge A\wedge A
\end{equation}
\noindent we find
\begin{align}
\mbox{Tr}\left\{\frac{CS(A) + CS(\bar{A})}{2}\right\} 
={}&
  \mbox{Tr}\left\{\omega \wedge d\omega +\frac{2}{3}\omega\wedge\omega\wedge\omega + \frac{e}{\ell^2}\wedge T \right\}
  \\
\mbox{Tr}\left\{\frac{CS(A) - CS(\bar{A})}{2}\right\}
={}&
\mbox{Tr}\left\{ \frac{2}{\ell}e\wedge R + \frac{2}{3\ell^{3}}e\wedge e \wedge e -\frac{1}{\ell}d(\omega \wedge e)\right\}\,,
\end{align}

\noindent where $R = R^{a}J_{a}$ and $T = T^{a}J_{a}\,$. A short calculation using $\det(e) = \sqrt{|g|}$ (where we assumed the positive orientation), $\epsilon_{abc}\, e^{a}\wedge R^{bc} = \sqrt{|g|}\mathcal{R}\,d^{3}x\,$ (where $\mathcal{R}$ denotes the Ricci scalar), and  $\epsilon_{abc}\, e^{a}\wedge e^{b}\wedge e^{c} = 3! \sqrt{|g|}\, d^{3}x\,$ reveals
\begin{align}
\mbox{Tr}\left\{\frac{2}{\ell}e\wedge R + \frac{2}{3\ell^{3}}e\wedge e \wedge e\right\}
={}&
\frac{y_{R}}{2\ell}\sqrt{|g|}\left(\mathcal{R} + \frac{2}{\ell^{2}}\right)d^{3}x\,,
\nonumber
\end{align}
\noindent where $y_{R}$ is a representation-dependent normalization constant defined through $\text{Tr}\left[J_{a}J_{b}\right] = (y_{R}/2)\eta_{ab}\,$. Taking $k  = \ell/(4G_{3})$, it follows that
\begin{align}\label{Chern Simons action}
 I ={}&
 \frac{k}{4\pi\,y_{R} }\int_{M} \mbox{Tr}\Bigl[CS(A) - CS(\bar{A})\Bigr]
\\
={}&
 \frac{1}{16\pi G_{3}}\left[\int_{M} d^{3}x\sqrt{|g|}\left(\mathcal{R} + \frac{2}{\ell^{2}}\right) -\int_{\partial M}\omega^{a}\wedge e_{a}\right],
 \nonumber
\end{align}
\noindent as claimed in the main text.

%%%%%%%%%%%%%%%%%%%%%%%%%%%%%%%%%%%%%%%%%%%%%%%%%%%%%%%%%%%%%%%%%%%%%%%%%%%%%%%%
\section{Conventions for the $sl(2)$ and $sl(3)$ algebras}\label{sec:algebras}
%%%%%%%%%%%%%%%%%%%%%%%%%%%%%%%%%%%%%%%%%%%%%%%%%%%%%%%%%%%%%%%%%%%%%%%%%%%%%%%%
Our convention fo the  $so(2,1)$ algebra is 
\begin{equation}
\left[J_{a},J_{b}\right] = \epsilon_{abc}J^{c}\,,
\end{equation}
\noindent where $J^{a} \equiv \eta^{ab}J_{b}$ and $\epsilon_{012} =-1\,$. The generators $\Lambda^{0},\Lambda^{\pm}$ defined through
\begin{equation}\label{def sl2 generators}
J_{0} = \frac{\Lambda^{+} + \Lambda^{-}}{2}\,,\quad J_{1} =  \frac{\Lambda^{+} -\Lambda^{-}}{2}\,,\quad J_{2} = \Lambda^{0}\,,
\end{equation}
\noindent then satisfy the $sl(2,\mathds{R}) \simeq so(2,1)$ algebra
\begin{equation*}
\left[\Lambda^{\pm},\Lambda^{0}\right] = \pm \Lambda^{\pm}\,,\,\,  \left[\Lambda^{+},\Lambda^{-}\right] = 2\Lambda^{0}\, .
\end{equation*}

\noindent The usual two-dimensional representation of $sl(2,\mathds{R})$ is in terms of matrices
\begin{equation}\label{2d representation of sl2r}
\Lambda^{0}= 
\frac{1}{2}\left(\begin{array}{cc}
1 & 0  \\ 
0 & -1
\end{array} 
\right),\qquad
\Lambda^{+}= 
\left(\begin{array}{cc}
0 & 0  \\ 
1 & 0 
\end{array} 
\right),
\qquad 
\Lambda^{-}= 
\left(\begin{array}{cc}
0 & -1  \\ 
0& 0 
\end{array} 
\right).
\end{equation}
\noindent The $so(2,1)$ generators in this representation are then $J_{0} = -i\sigma^{y}/2$, $J_{1}=\sigma^{x}/2$, $J_{2}=\sigma^{z}/2$, where the $\sigma$'s are the Pauli matrices.

Similarly, we can parameterize the $sl(3,\mathds{R})$ algebra in terms of generators $L_{0}$, $L_{\pm 1}$ which span an $sl(2,\mathds{R})$ subalgebra, augmented by five $W_{j}$ generators ($j = -2, -1,0,1,2$) forming a spin-$2$ multiplet under the triplet $L_0$, $L_{\pm 1}\,$, with commutation relations
\begin{align*}%\label{sl3R algebra}
\left[L_{j},L_{k}\right] 
&=
 (j-k)L_{j+k}
\nonumber\\
\left[L_{j}, W_{m}\right] &=
 (2j - m)W_{j+m}
\\
\left[W_{m},W_{n}\right] &=
 -\frac{1}{3}\left(m-n\right)\left(2m^{2} + 2n^{2} -mn-8\right)L_{m+n}\, .
\end{align*}

\noindent With this parameterization the principal and diagonal embeddings correspond to identifying the $sl(2,\mathds{R})$ generators as
\begin{align}
\text{principal embedding:} \qquad \{\Lambda^0,\Lambda^{\pm}\} =& \left\{L_{0}, L_{\pm 1}\right\}
\\
\text{diagonal embedding:} \qquad \{\hat{\Lambda}^0,\hat{\Lambda}^{\pm}\} =& \left\{\frac{1}{2}L_{0}, \pm\frac{1}{4}W_{\pm 2}\right\}
\end{align}

%%%%%%%%%%%%%%%%%%%%%%%%%%%%%%%%%%%%%%%%%%%%%%%%%%%%%%%%%%%%%%%%%%%%%%%%%%%%%%%%
\section{Some $sl(N)$ representation theory\label{sec:reptheory}}
%%%%%%%%%%%%%%%%%%%%%%%%%%%%%%%%%%%%%%%%%%%%%%%%%%%%%%%%%%%%%%%%%%%%%%%%%%%%%%%%
Here we collect some useful facts from the representation theory of $sl(N)\,$. The $sl(N)$ algebra is a semi-simple algebra of rank $N-1\,$. In order to write down its weights and roots, we will first construct a convenient basis for the $(N-1)$-dimensional weight space (i.e. the vector space dual to the Cartan subalgebra). To this end, let $\hat{e}_{i}$ with $i=1,\ldots, N$ denote the orthonormal basis of $\mathds{R}^{N}$, and define $\hat{\gamma} = \sum_{i=1}^{N}\hat{e}_{i}\,$. We then define vectors $e_{i}$ by projecting the $\hat{e}_{i}$ onto a plane orthogonal to $\hat{\gamma}$: 
\begin{equation}
e_{i} = \hat{e}_{i} - \frac{\hat{\gamma}}{N}\,.
\end{equation}

\noindent Notice the $e_{i}$ satisfy $\sum_{i=1}^{N}e_{i}=0$ and can be thought of as (linearly dependent) vectors in weight space. Their inner products are given by
\begin{equation}
\langle e_{i},e_{j}\rangle = \delta_{ij} - \frac{1}{N}\,.
\end{equation}

\noindent The positive roots can then be written as
\begin{equation}
\alpha_{ij} = e_{i}-e_{j}\,,\quad i< j \qquad \Rightarrow \qquad \text{Number of positive roots } =\frac{N(N-1)}{2}\,.
\end{equation}

\noindent The $N-1$ simple roots correspond to $\alpha_{i} \equiv \alpha_{i\, i+1} = e_{i} - e_{i+1}$ ($i=1,\ldots, N-1$). The Cartan matrix is
\begin{equation}
C_{ij} = 2\frac{\langle \alpha_i,\alpha_j\rangle}{\langle \alpha_j,\alpha_j\rangle} = \langle \alpha_i,\alpha_j\rangle = 
\left\{
\begin{array}{rl}
2 & \text{if } i=j\\ 
-1 & \text{if } i=j \pm 1\\ 
0 & \text{otherwise}
\end{array} \right.
\end{equation}

\noindent and the Killing form $(\cdot,\cdot)$ is such that
\begin{equation}
 \langle \alpha ,\beta \rangle = \left(H_{\alpha},H_{\beta}\right) =  \frac{1}{2N}\text{Tr}_{Adj}\left[H_{\alpha}H_{\beta}\right].
\end{equation}

\noindent With these conventions, the fundamental weights $\omega_{i}$ ($i=1,\ldots , N-1$) are
\begin{equation}
\omega_{i} = \sum_{j =1}^{i}e_{j}\,.
\end{equation}

\noindent An important object is the so-called \textit{Weyl vector} $\vec{\rho}$, defined as the sum of the fundamental weights (with unit coefficient):
\begin{equation}
\vec{\rho} = \sum_{i=1}^{N-1}\omega_{i}
\end{equation}
\noindent Using $\sum_{i=1}^{N}e_{i}=0$ we can easily write it as
\begin{equation}
\vec{\rho} = \frac{(N-1)}{2}e_{1} + \frac{(N-3)}{2}e_{2} +\ldots + \frac{(N-2i+1)}{2}e_{i}+\ldots  - \frac{(N-3)}{2}e_{N-1}-\frac{(N-1)}{2}e_{N}\,.
\end{equation}

\noindent We see that regarded as a vector in $\mathds{R}^{N}$, i.e. in terms of the orthonormal basis $\hat{e}_{i}\,$, the Weyl vector has components $\vec{\rho} = \left(\frac{N-1}{2},\frac{N-3}{2},\ldots , -\frac{N-3}{2} ,-\frac{N-1}{2}\right)\,$. An important observation used in the main text is that, in the principal embedding, these are precisely the entries of the (diagonal) matrix $\Lambda^0$ in the $N$-dimensional (defining) representation. This can be also understood as follows. First, since $\Lambda^0$ belongs to the Cartan subalgebra, we can define its dual vector $\lambda_0$ in weight space via the usual isomorphism provided by the Killing form. Since the fundamental weights span the dual space, we can write $\lambda_0 = \sum a_{i}\omega_i$ for some coefficients $a_{i}\,$. On the other hand, if $\alpha_{j}$ denote the simple roots, in the principal embedding one has \cite{deBoer:1992sy} 
\begin{equation}
\text{principal embedding:}\qquad \Lambda^+ = \sum_{j=1}^{N-1}c_{j}E_{-\alpha_{j}}\,,\qquad \text{all }c_{j} \neq 0
\end{equation}
\noindent Then, using the commutation relations we find
\begin{align}
\Lambda^+ = - \left[\Lambda^0,\Lambda^+\right] =-\sum_{j}c_{j} \left[\Lambda^0,E_{-\alpha_{j}}\right]
 ={}&
 \sum_{j}c_{j} \langle \lambda_0, \alpha_{j}\rangle E_{-\alpha_{j}} 
 \nonumber\\
 ={}&
 \sum_{j} \sum_{i} c_{j} a_{i} \langle \omega_{i}, \alpha_{j}\rangle E_{-\alpha_{j}} \,.
\end{align}

\noindent Now, by definition the fundamental weights satisfy
\begin{equation}
\langle \omega_{i}, \alpha_{j}\rangle = \frac{\langle \alpha_i,\alpha_i\rangle}{2}\delta_{ij} = \delta_{ij}\,,
\end{equation}

\noindent and we conclude
\begin{equation}
\sum_{j=1}^{N-1}c_{j}E_{-\alpha_{j}} = \Lambda^+ =\sum_{j=1}^{N-1} a_{j} c_{j} E_{-\alpha_{j}} \quad \Leftrightarrow \quad a_{j} =1 \quad \forall j
\end{equation}

\noindent this is,
\begin{equation}
\text{principal embedding:}\qquad \lambda_0 = \sum_{i=1}^{N-1}\omega_{i} = \vec{\rho}
\end{equation}

\noindent In particular, in the principal embedding we have
\begin{equation}
 \text{Tr}_{N}\left[\Lambda^0\Lambda^0\right] = \langle \vec{\rho} ,\vec{\rho} \rangle = \frac{N(N^{2}-1)}{12}\,.
\end{equation}

\providecommand{\href}[2]{#2}\begingroup\raggedright\endgroup

\end{document}